\documentclass[twocolumn]{aastex61}

\newcommand{\msun}{M$_{\odot}$}
\newcommand{\myr}{M$_\odot$~yr$^{-1}$}

\newcommand{\ha}{H$\alpha$}
\newcommand{\lha}{L$_{H\alpha}$}
\newcommand{\rha}{r$_{H\alpha}$}
\newcommand{\kms}{km\,s$^{-1}$}

\newcommand{\sfrd}{$\Sigma_{SFR}$}
\newcommand{\fgas}{$f_{gas}$}
\newcommand{\hii}{H {\sc II} }
\newcommand{\oii}{O{\sc II}}

\usepackage{amsmath}

\shorttitle{Star-Forming Clump Scaling Relations}
\shortauthors{Cosens et al.}

\begin{document}

\title{Size-luminosity scaling relations of local and distant star forming regions}

\correspondingauthor{Maren Cosens}
\email{mcosens@ucsd.edu}

\author[0000-0002-2248-6107]{Maren Cosens}
\affil{Physics Department, University of California, San Diego, 9500 Gilman Drive, La Jolla, CA 92093 USA} 
\affil{Center for Astrophysics and Space Sciences, University of California, San Diego, 9500 Gilman Drive, La Jolla, CA 92093 USA}
\author[0000-0003-1034-8054]{Shelley A. Wright}
\affil{Physics Department, University of California, San Diego, 9500 Gilman Drive, La Jolla, CA 92093 USA} 
\affil{Center for Astrophysics and Space Sciences, University of California, San Diego, 9500 Gilman Drive, La Jolla, CA 92093 USA}
\author{Etsuko Mieda}
\affil{Subaru Telescope, National Astronomical Observatory of Japan, 650 North A'ohoku Place, Hilo, HI 96720 USA}
\author{Norman Murray}
\affil{Canadian Institute for Theoretical Astrophysics, University of Toronto, 60 St. George Street, Toronto, ON M5S 3H8, Canada}
\affil{Canada Research Chair in Astrophysics}
\author[0000-0003-3498-2973]{Lee Armus}
\affil{Spitzer Science Center, California Institute of Technology, 1200 E. California Blvd., Pasadena, CA 91125 USA}
\author[0000-0001-9554-6062]{Tuan Do}
\affil{UCLA Galactic Center Group, Department of Physics \& Astronomy, University of California, Los Angeles, Los Angeles, CA 90095, USA}
\author{James E. Larkin}
\affil{Department of Physics \& Astronomy, University of California, Los Angeles, Los Angeles, CA 90095, USA}
\author{Kirsten Larson}
\affil{Spitzer Science Center, California Institute of Technology, 1200 E. California Blvd., Pasadena, CA 91125 USA}
\author{Gregory Martinez}
\affil{UCLA Galactic Center Group, Department of Physics \& Astronomy, University of California, Los Angeles, Los Angeles, CA 90095, USA}
\author{Gregory Walth}
\affil{Center for Astrophysics and Space Sciences, University of California, San Diego, 9500 Gilman Drive, La Jolla, CA 92093 USA}
\author[0000-0002-0710-3729]{Andrey Vayner}
\affil{Physics Department, University of California, San Diego, 9500 Gilman Drive, La Jolla, CA 92093 USA} 
\affil{Center for Astrophysics and Space Sciences, University of California, San Diego, 9500 Gilman Drive, La Jolla, CA 92093 USA} 

\received{July 13, 2018}
\revised{October 19, 2018}
\accepted{October 23, 2018}
\submitjournal{The Astrophysical Journal}

\begin{abstract}
We investigate star forming scaling relations using Bayesian inference on a comprehensive data sample of low- (z$<$0.1) and high-redshift (1$<$z$<$5) star forming regions. This full data set spans a wide range of host galaxy stellar mass ($M_{*} \sim10^6-10^{11}$ \msun) and clump star formation rates (SFR $ \sim10^{-5}-10^2$ \myr). We fit the power-law relationship between the size (\rha) and luminosity (\lha) of the star forming clumps using the Bayesian statistical modeling tool Stan that makes use of Markov Chain Monte Carlo (MCMC) sampling techniques. Trends in the scaling relationship are explored for the full sample and subsets based on redshift and selection effects between samples. In our investigation we find no evidence of redshift evolution of the size-luminosity scaling relationship, nor a difference in slope between lensed and unlensed data. There is evidence of a break in the scaling relationship between high and low star formation rate surface density (\sfrd) clumps. The size-luminosity power law fit results are \lha$\sim$ \rha$^{2.8}$ and \lha$\sim$ \rha$^{1.7}$ for low and high \sfrd \, clumps, respectively. We present a model where star forming clumps form at locations of gravitational instability and produce an ionized region represented by the Str\"{o}mgren radius. A radius smaller than the scale height of the disk results in a scaling relationship of $L \propto r^3$ (high \sfrd \, clumps), and a scaling of $L \propto r^2$ (low \sfrd \, clumps) if the radius is larger than the disk scale height.

\end{abstract}
\keywords{galaxies: high redshift -- galaxies: star formation -- galaxies: morphologies -- methods: statistical -- methods: data analysis -- techniques: imaging spectroscopy}

\section{Introduction}

Understanding the star formation properties in high-redshift galaxies is crucial for understanding galactic formation and evolution. Star formation rates at high-redshift (z$\sim$2) are an order of magnitude higher than at z$\sim$0 \citep{hop04, hop06, mad14}, indicating that the majority of stellar mass and galactic substructure are established at early times. Rest frame UV \textit{Hubble Space Telescope} (\textit{HST}) imaging surveys implied star formation occurred in irregular morphologies \citep[e.g.,][]{elm04a, elm04b, law07a}, while ground-based spectroscopic surveys confirmed the large global star formation properties of high-redshift galaxies \citep{shap03, law07a}. Yet these early surveys were unable to resolve individual star forming regions (``clumps") to study their internal kinematics and sizes. Studying the properities of individual high-redshift star forming clumps is imperative for comparing their properties to that of local \hii regions and starburst regions, and for understanding their star formation mechanisms.

Integral field spectrographs (IFS) have been revolutionary for studying the resolved morphologies and kinematics of high redshift galaxies \citep{glaz13}. Using an IFS in combination with Adaptive Optics (AO) yields superb spatial resolutions, down to $\sim800$ pc at $z\sim1$. This has allowed for detailed ionized gas kinematic studies of high-redshift galaxies and their individual clumps \citep{fors06, fors09, fors11, genz06, genz08, genz11, law07b,law09, wright07, wright09, shap08, ep09, ep12, swin09, swin12a, swin12b, jon10, man11, liv15, wisn12, wisn15, new13, buit14, sto14, sto16, lee16, mied16, mol17}. The kinematics of these galaxies have shown large turbulent disks that have high velocity dispersions ($>>$10 \kms). These high-redshift disks have had their Toomre parameter, Q, measured to be less than 1 \citep{toom64}, and therefore gravitational instability \citep{elm08, genz11} may cause disk fragmentation and clump formation \citep[e.g.][]{bour07, elm08, man14}.

In a large HST imaging-survey, \citet{guo15} finds that the majority of high-redshift galaxies contain one or more off-center clumps, where the number of clumps per galaxy is decreasing with redshift to $z\approx0.5$. These clumps are larger than local Giant Molecular Clouds (GMCs) and \hii regions with size scales on the order of $\sim 1-3$ kpc, and only a small number of clumps in each galaxy as opposed to hundreds of GMCs and \hii regions in local galaxies. One interpretation is that these massive clumps coalesce to form or grow the bulge of their host galaxy, spiraling towards the center due to the effects of dynamical friction \citep{bour07, elm08}. The migration of massive clumps towards the center of the host galaxy is thought to occur on timescales of $\sim 2-3$ orbital times \citep{dek09, cev12, bour14, man14, man17}. This process would then lead to the exponential disk structure we typically see in local spiral galaxies \citep{bour07}. 

The ability to measure resolved clump properties provides insight into the physical processes driving high-redshift clump formation, and how these systems evolve into local galaxies. To explore the driving formation mechanisms, star formation scaling relations of high-redshift clumps are often compared to local analogs like \hii regions. The relationships between clump size, luminosity (usually in \ha; \lha), and velocity dispersion have been investigated in various studies with differing results \citep{genz11, wisn12, liv12, liv15, mied16}. In comparison to local \hii regions, \citet{liv12,liv15} (the latter including data from \citet{jon10}) find there is an offset to higher luminosities in their lensed, high-redshift clumps. However, both \citet{wisn12} and \citet{mied16} find that the power-law relating clump size and luminosity for unlensed high-redshift samples extend well to local \hii regions, with \citet{wisn12} finding the relationship $L_{H\alpha} \propto r^{2.72\pm0.04}$ when including local \hii and giant \hii regions. In order to determine whether these scaling relationship differences are due to redshift evolution, selection biases between studies, and/or intrinsic scatter requires additional local and high-redshift investigations.

An important consideration for studying high-redshift scaling relations is which local analogs to use as a comparison sample. Often \hii regions like those found in the SINGS survey \citep{kenn03} are used as these comparative local analogs. However, high-redshift star forming clumps are sometimes found to be orders of magnitude more luminous than local \hii regions \citep{swin09}, and may in fact be scaled up versions of more extreme giant \hii regions such as 30 Doradus \citep{swin09, jon10, wisn12}. The DYNAMO survey \citep{fish16} provides another set of local clump analogs in turbulent galaxies that have similar properties to high-redshift clumps. Within the Milky Way there are distinctions between star forming regions based on size-scale, where GMCs are 1 to 2 orders of magnitude smaller than  Molecular Cloud Complexes (MCCs). \citet{ngu16} investigate a power-law break in varying star formation laws based on the differences between GMCs and MCCs that indicate MCCs may provide another analog to the high-redshift clumps.

We gathered a comprehensive data set from the literature to form a robust comparative sample in Section \ref{sec:data} to investigate possible causes of variation in the scaling relations between different samples. In Section \ref{sec:statistics} we discuss the Markov Chain Monte Carlo (MCMC) method developed to fit a power-law to clump sizes and luminosities. In Section \ref{sec:results}, we present the results of this fitting method for a range of data subsets to investigate the clump size-luminosity scaling relationship. We apply a broken power-law fit to this relationship based on the star formation rate surface density, as presented in Section \ref{sec:SFRDbreak}. The possible effects of beam smearing on the measured clump properties and scaling relations are explored in Section \ref{sec:blurring}. We divide the data into various sub-samples to investigate potential redshift evolution in Section \ref{sec:redshift}; and dependence on clump velocity dispersion and host galaxy gas fraction in Section \ref{sec:3D}. Lastly, in Section \ref{sec:discussion} we discuss two potential theoretical models that may explain the size-luminosity relationships measured. We present a new model that re-scales the Str\"{o}mgren sphere in context to the galaxy disk size with large star forming clumps. We further discuss any observed biases and selection effects that could influence the fitting to the star forming clump scaling relationship. In Section \ref{sec:conclusion} we summarize our results. Throughout this paper we use the concordance cosmology with $\rm H_0 = 67.8 km \ s^{-1} \ Mpc^{-1}$, $\Omega_M = 0.306$, and $\Omega_{\Lambda}= 0.692$ \citep{planck14}.

\section{Data Sample}\label{sec:data}

\begin{figure}[h]
\gridline{\fig{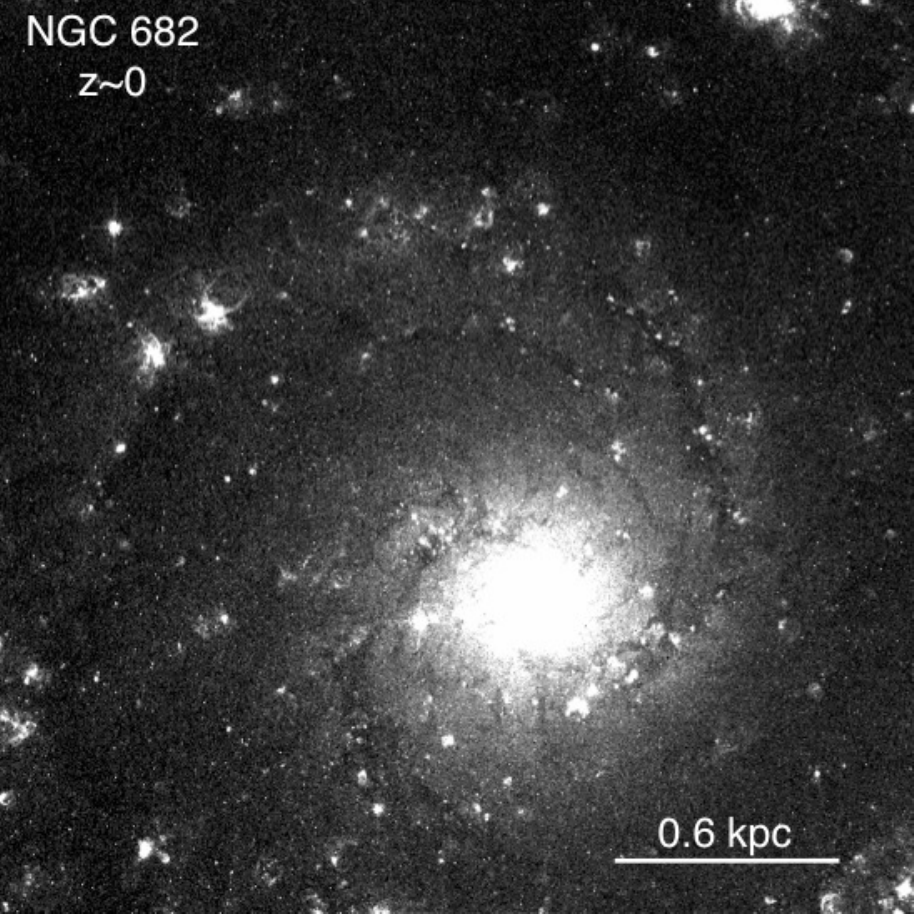}{0.22\textwidth}{(a): local, low \sfrd}
            \fig{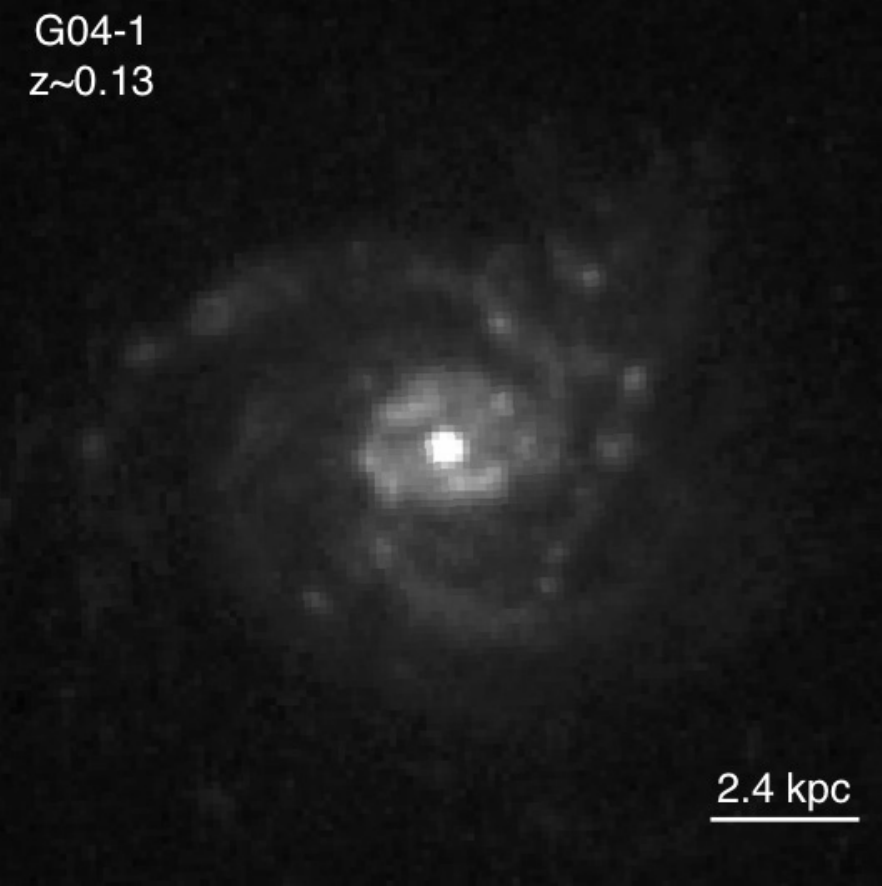}{0.22\textwidth}{(b): local, high \sfrd}}
\gridline{\fig{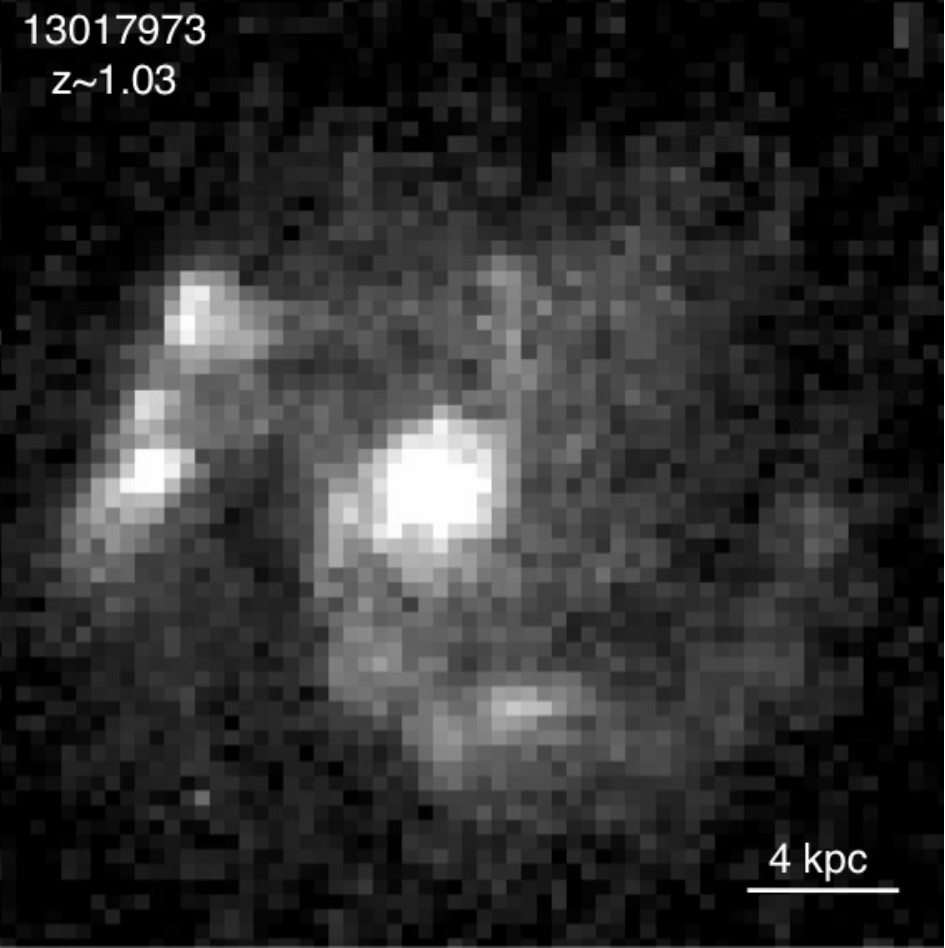}{0.22\textwidth}{(c): high-redshift, unlensed}
            \fig{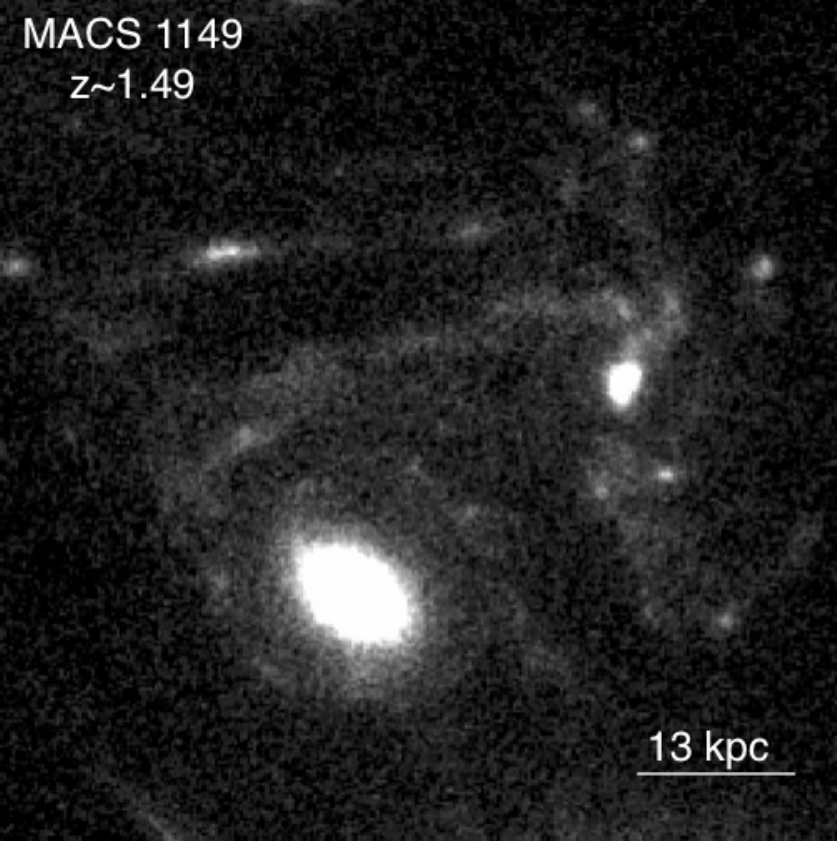}{0.22\textwidth}{(d): high-redshift, lensed}}
\caption{HST images from archival ACS data of galaxies representative of the variation in morphology within the full sample investigated here. (a): local SINGS galaxy NGC 628 included in the sample of $z\approx0$ \hii regions \protect\citep[][data set \#10 in Table \ref{tbl:data_sets}]{kenn03} taken with ACS F658W filter. (b): turbulent local galaxy from the DYNAMO sample \protect\citep[][data set \#11]{fish16} taken in the ACS/WFC1-IRAMP FR716N filter.  (c): $z\sim1$ unlensed galaxy from the IROCKS sample \protect\citep[][data set \#9]{mied16} imaged in the F814W filter with ACS. (d): $z\sim1.5$ lensed galaxy MACS 1149 \protect\citep[][data set \#3]{liv15} taken with the ACS F814W filter. Scale is at the redshift of MACS 1149 without taking into account the lensing effects which cause the spatial resolution to vary across the galaxy. \label{fig:HST}}
\end{figure}

Data of star forming clumps from both high and low redshift ($z\sim 0.6-5; z\sim 0-0.1$) galaxies measured and detected in different ways were gathered from the literature to form a comprehensive sample of the known data \citep{swin09, jon10, liv12, liv15, wal18, genz11, wisn12, freun13, mied16, kenn03, gal83, ars88, bas06, roz06, mon07, fish16, ngu16}.  This sample is detailed in Table \ref{tbl:data_sets} and includes lensed \citep{swin09, jon10, liv12, liv15, wal18} and unlensed \citep{genz11, wisn12, freun13, mied16} high-redshift galaxies, as well as a wide range of sizes and star formation rate densities in the local analogs \citep{kenn03, gal83, ars88, bas06, roz06, mon07, fish16, ngu16}. Figure \ref{fig:HST} illustrates the differences in the morphologies of these galaxy populations via a comparison of HST images of representative objects.

\begin{figure*}
\gridline{\fig{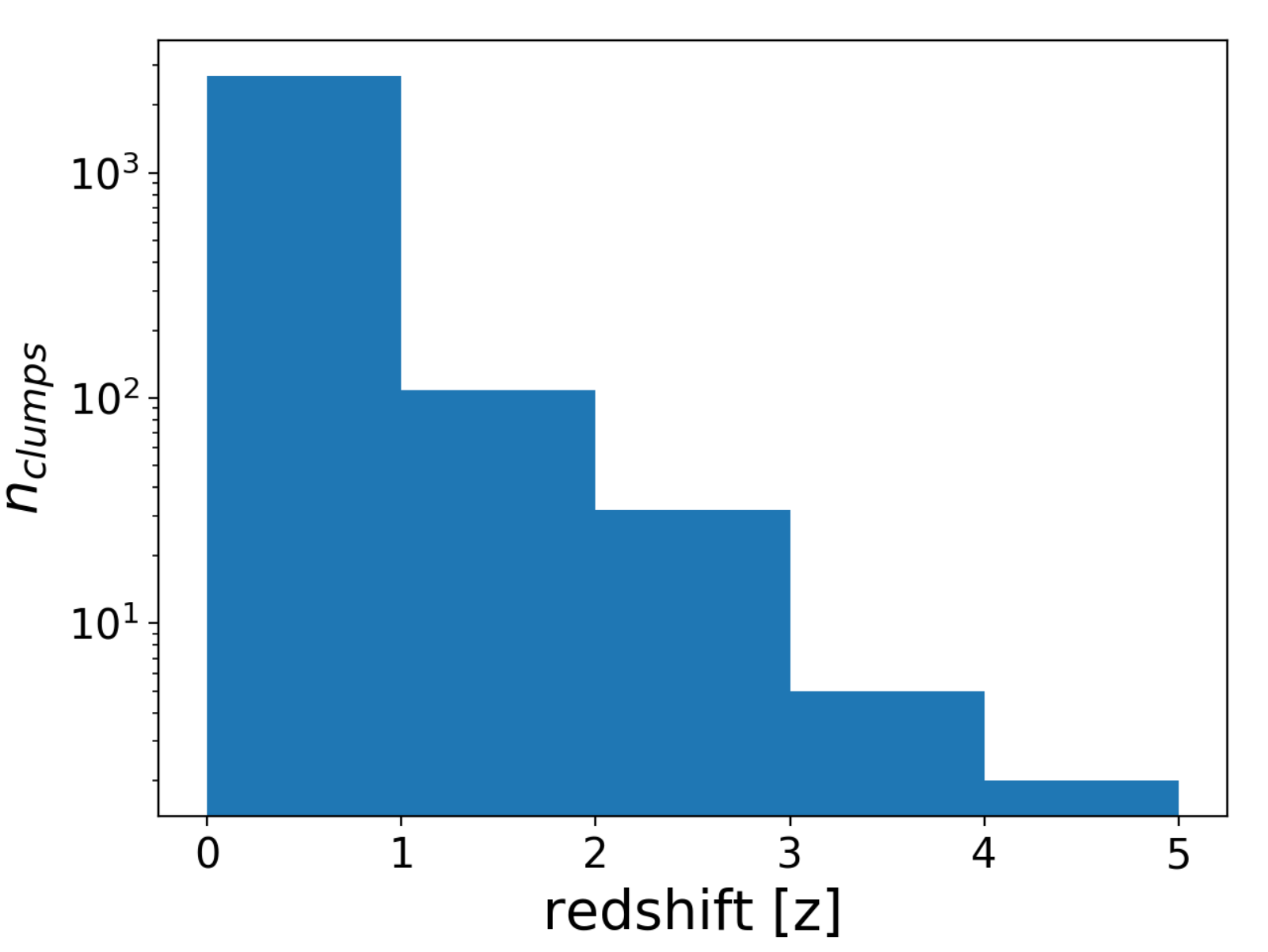}{0.32\textwidth}{(a)}
            \fig{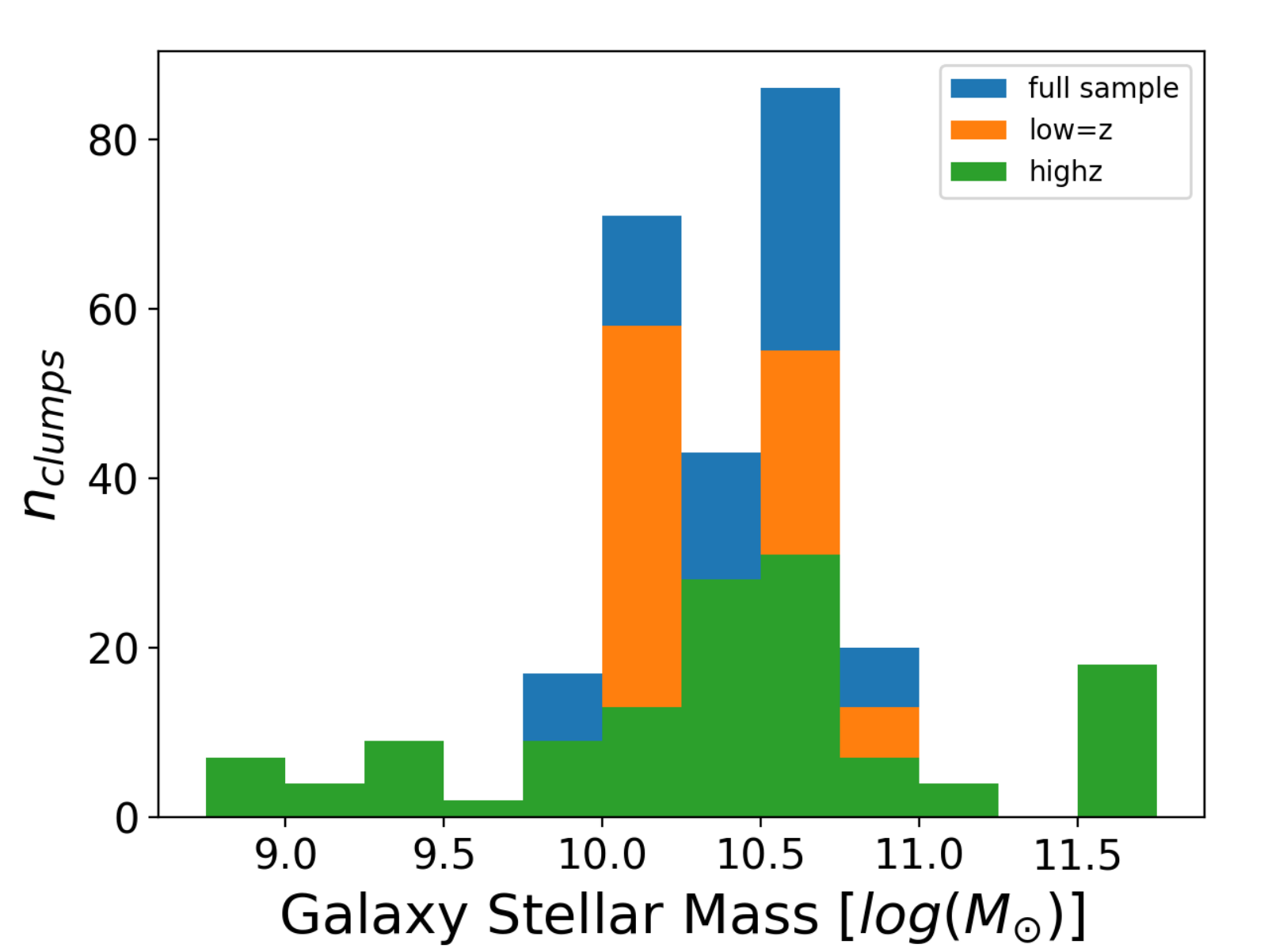}{0.32\textwidth}{(b)}
            \fig{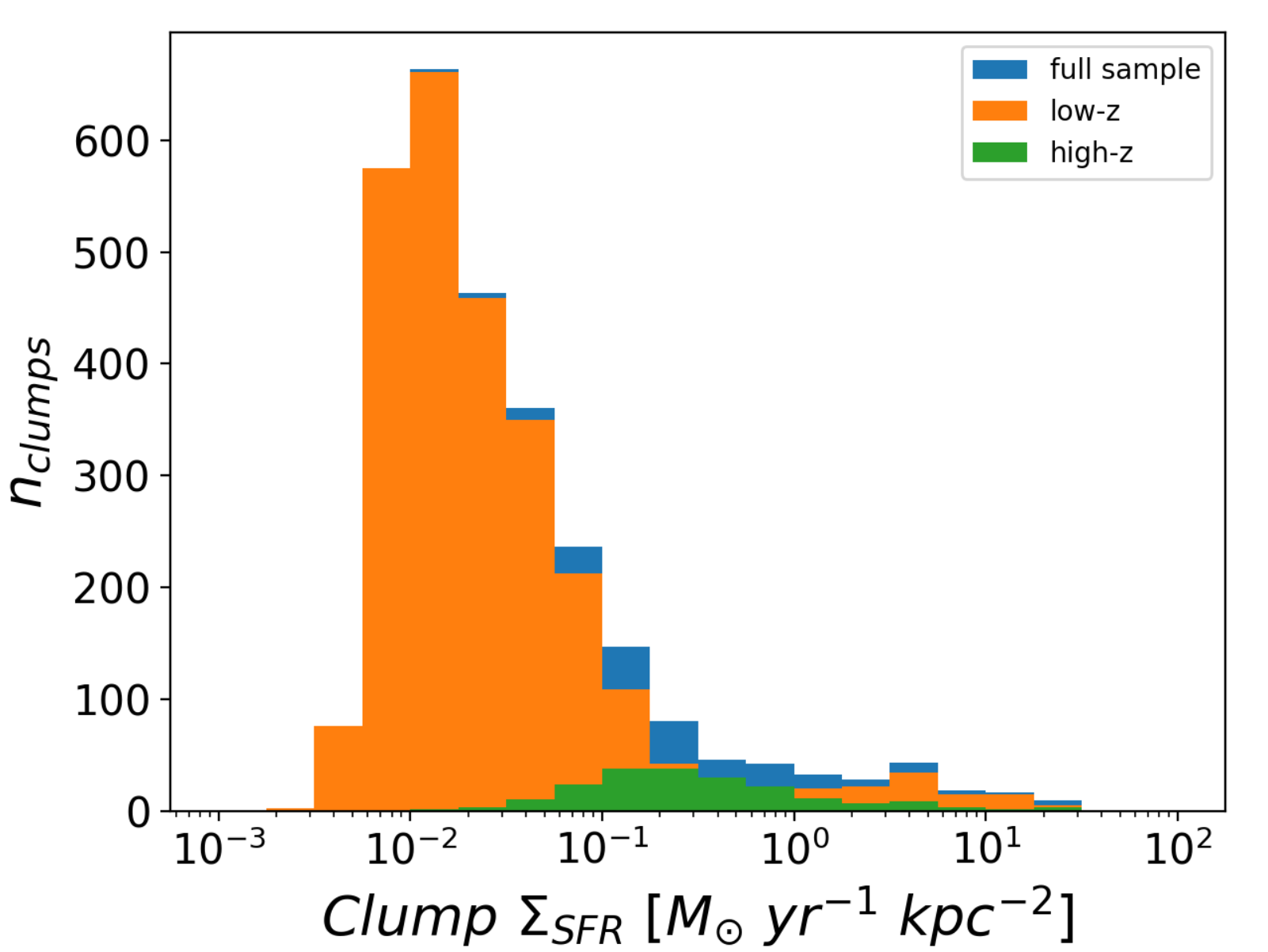}{0.32\textwidth}{(c)}}
\caption{Histograms showing the distribution of clump properties for the full sample. (a): redshift of each clump. The abundance of local samples and difficulty of observing higher redshift galaxies leads to the bias towards low redshift seen here. (b): Stellar mass of the host galaxy for each clump used. Values of host galaxy stellar mass were not reported for data set numbers 4, 5, \& 10 as designated in Table \ref{tbl:data_sets}. (c): \sfrd \, of each clump. The bias towards lower \sfrd \, comes from the high numbers of local \hii regions with lower \sfrd. \label{fig:histograms}}
\end{figure*}

The majority of high-redshift samples make use of IFS systems for investigating the morphological and kinematic properties of the star forming clumps. This allows for detailed study of the kinematics of the galaxy at improved spatial resolution when coupled with AO. The range of properties spanned by the full sample is shown in the histograms of Figure \ref{fig:histograms}. The set of high-redshift unlensed galaxies ($z\sim 1-2$) have an average stellar mass of $\sim 10^{11}$ \msun \, and an average spatial resolution of 2000 pc ($\sim 0.6"$). The high-redshift lensed galaxies ($z\sim 0.6-5$) tend to have a lower overall stellar mass ($\sim 10^8$ \msun), but better spatial resolution (avg $\sim 300$ pc; $\sim 0.05"$) than the unlensed galaxies.  The difference in the sampling of the lensed and unlensed surveys leads to the bimodal appearance of the histogram of host galaxy stellar mass (Figure \ref{fig:histograms}b). The various local analogs span a wide range of total stellar masses ($\sim 10^{6} - 10^{12}$ \msun) with spatial resolution similar to or slightly better than the high-redshift lensed sample.  This wide range of local analogs provides a robust comparison to the varied high-redshift clumps observed.

\clearpage
\movetabledown=1.25in
\begin{rotatetable*}
\begin{deluxetable*}{lccccccc}
\tabletypesize{\tiny}
\tablecaption{Data Samples: Observational Properties of High and Low Redshift Star Forming Clumps \label{tbl:data_sets}}
\tablehead{
\colhead{Study(ies)} & \colhead{Redshift} & \colhead{Instrument} & \colhead{Spatial Resolution} & \colhead{Spatial Resolution} & \colhead{Lensed/Unlensed} & \colhead{Galaxy $M_{*}$ Range} & \colhead{Galaxies} \\
\colhead{\#, References} & \colhead{(z)} & \colhead{} & \colhead{(arcsec)} & \colhead{(pc)} & \colhead{} & \colhead{($M_{\sun}$)} & \colhead{(\#)}}
\startdata
1, \citet{swin09} & 4.9 & Gemini/NIFS & 0.06" & 320 & lensed & $7\pm2\times10^8$ & 1 \\
\tableline
2, \citet{jon10} & $\sim$1.7-3.1 & Keck/OSIRIS & $0.01"^{a}$ & $\sim 100$ & lensed & $10^{9.7-10.3}$ (dynamical) & 6 \\
\tableline
3, \citet{liv15} & & VLT/SINFONI, & & & & & 10 \\
& $\sim$1-4 & Keck/OSIRIS, & $\sim0.04"-0.08"^{a}$ & 40-700 & lensed & $4\times10^8 - 6\times10^8$ & 1 \\
& & Gemini/NIFS & & & & & 1\\
\tableline
4, \citet{liv12} & $\sim$1-1.5 & WFC3 & 0.05" & $\sim70-600$ & lensed & \nodata & 8 \\
\tableline
5, \citet{wal18} & 0.61 & HST/ACS$^{b}$ \& WFC3$^{b,c}$ & $0.01"/0.03"^{a}$ & 90/240 & lensed & $2.6\times10^{10}$ & 1 \\
& & Magellan/LDSS-3$^{e}$ \& MMIRS$^{e}$ & & & & \\
\tableline
6, \citet{genz11} & $\sim$2 & VLT/SINFONI & $\sim0.2"$ & $\sim1700^{a}$ & unlensed & $\sim10^{10.6}$ & 5 \\
\tableline
7, \citet{wisn12} & $\sim$1.3 & Keck/OSIRIS & $\sim0.1"$ & $\sim520-840$ & unlensed & $\sim10^{11}$ & 3 \\
\tableline
8, \citet{freun13} & $\sim$1.2 & IRAM \& Keck/DEEP2 & $0.6"-1.9"^{f}$ & $\sim8000$ & unlensed & $\sim10^{11}$ & 4 \\
\tableline
9, \citet{mied16} & $\sim$1 & Keck/OSIRIS & $\sim0.1"$ & 800 & unlensed & $10^{9.6-11.2}$ & 7 \\
\tableline
\ \ \ \ \citet{kenn03}$^{d}$ & & KPNO \& CTIO & 1"-3" & 40-325$^{a}$ &  & \nodata  & 7 \\
\ \ \ \ \citet{gal83}$^{e}$ & & Kitt Peak video camera system & \nodata & 200$^{g}$ &  &  \nodata & 10 \\
10, \citet{ars88}$^{e}$ &  & various & >4" & >100$^{a}$&  &  \nodata &  20\\
\ \ \ \ \citet{bas06}$^{e}$ & $\sim$0 & VLT-VIMOS & 0.66" & $\sim50^{a}$ & unlensed & \nodata & 2 \\
\ \ \ \ \citet{roz06}$^{e}$ &  & OAN-SPM \& William Herschel Telescope & $1.5"-1.6"$ & $\sim50-160^{a}$ & &  &  10 \\
\ \ \ \ \citet{mon07}$^{e}$ &  & INTEGRAL/WYFFOS \& WFPC2 & \nodata & \nodata &  & $\sim2\times10^6 - 7\times10^8$ & 5\\
\tableline
11, \citet{fish16} & $\sim$0.1 & HST/WFC & $\sim0.05"$ & $\sim100$ & unlensed & $1-9\times10^{10}$ & 10 \\
\tableline
12, \citet{ngu16}$^{h}$ & Milky Way & CfA Suvey & 8.8' & $\sim 15$ & unlensed & $\sim10^{10}$ & 1 \\
\enddata
\tablecomments{$^{a}$: when resolution was only given in either pc or arcsec it was converted to the other units based on the cosmology used here. $^{b}$: used for measurement of region size. $^{c}$: used for measurement of flux. $^{d}$: normal \hii regions; re-analyzed by \protect\citet{wisn12}. $^{e}$: Giant \hii regions; corrections applied by \protect\citet{wisn12}. $^{f}$: clump sizes for the \protect\citet{freun13} sample are derived from IRAM CO luminosity with FWHM ranges given here; SFRs are derived from DEEP2 spectra using a 1" slit. $^{g}$: \ha \, flux measured within a fixed aperture diameter of 200pc \protect\citep{gal83}. $^{h}$: Molecular Cloud Complexes (MCC's).}
\end{deluxetable*}
\end{rotatetable*}

\section{Analysis: Bayesian Inference}\label{sec:statistics}
Previous investigations of high-redshift clumps have employed least-squares fitting to determine clump scaling relations (i.e. \citet{wisn12, mied16}). However, standard weighted least-squares relies on many assumptions about the inputs; to truly be reliable there are strict constraints that are often not really the case for the data \citep[]{hogg10}. These constraints are that one dimension has negligible uncertainties and the uncertainties in the other dimension are Gaussian with a known variance.

Most often there will be non-negligible uncertainties in both dimensions, and these uncertainties are  not always Gaussian. An approximation to meeting the constraints above would be to propagate the uncertainties of both dimensions to an overall uncertainty for each data point, but this is only an approximation and therefore not as reliable as including the uncertainties on their respective dimension. This approximated uncertainty also may not be Gaussian, violating the second constraint.

Another possible method for determining the scaling relations is to employ Bayesian inference along with Markov Chain Monte Carlo (MCMC) sampling. Bayesian analysis maps the posterior distribution, but the models can be complex and it is extremely difficult to work with this distribution directly. MCMC methods provide a way to sample the distribution and produce well defined statistical estimates of model parameters \citep[]{tier94}.

Bayes' Theorem in its most basic form is stated as follows:
\begin{equation}
    \mathcal{P}(A | B) = \frac{\mathcal{P}(B | A) \mathcal{P}(A)}{\mathcal{P}(B)}
\end{equation}
$\mathcal{P}(A | B)$ is the likelihood of event A given that B is true (conditional probability); $\mathcal{P}(B | A)$ is the likelihood of event B given A is true; and $\mathcal{P}(A)$ and $\mathcal{P}(B)$  are the likelihood of observing A and B independently (marginal probability). In Bayesian inference $\mathcal{P}(A)$ is known as the ``prior" and $\mathcal{P}(A | B)$ as the ``posterior".
For a set of data points or events, $\mathcal{P}(B)=\sum_{j=0}^{N}\mathcal{P}(B | A_j) \mathcal{P}(A_j)$ and Bayes Theorem becomes:
\begin{equation}
    \mathcal{P}(A_i | B) = \frac{\mathcal{P}(B | A_i) \mathcal{P}(A_i)}{\sum_{j=0}^{N}\mathcal{P}(B | A_j) \mathcal{P}(A_j)}
\end{equation}

By using Bayesian inference we are able to easily account for intrinsic scatter in the relationship as well as measured uncertainties in \textbf{both} dimensions without approximating to an overall uncertainty.  We are also able to include previously known information about the data and relationship through the priors \citep[]{ber85}. Priors essentially define the domain of the parameters we are trying to determine in the fitting. How we choose these priors is informed both theoretically and empirically by previous data and fitting. Additionally, we can not only determine an estimate for a model parameter, but also an uncertainty for that estimate, meaning that we can determine the best fit and have a well defined uncertainty for that model. This comes from the fact that Bayesian analysis produces a distribution for the unknown model parameters, the posteriors \citep[]{ber85}. 

Given the advantages of Bayesian inference as well as the shortcomings of a traditional least squares fit it seems prudent to employ Bayesian inference to investigate the clump scaling relations.

\subsection{Fitting Data Using PyStan} \label{sec:PyStan}
Data from star forming clumps in local and high-redshift galaxies were fit using PyStan, the Python interface to Stan, a tool for Bayesian statistical modeling using MCMC sampling techniques \citep{stan17}. This MCMC script was run including measurement uncertainty on both clump luminosity and size. One difficulty in this fitting process is determining how best to incorporate uncertainties since each study being included determines their uncertainties differently. Some studies have very large uncertainties while others are very small or not calculated at all. Even within similar studies (i.e. lensed vs unlensed or similar instrument and redshift) the size of the uncertainties is not consistent. For example, in the unlensed sample, \citet{wisn12} has uncertainties on clump radius and luminosity, \citet{mied16} only has uncertainty for luminosity\footnote{\citet{mied16} determined that the definition used for a clump had a larger impact on the uncertainty of the radius than the measurement error itself.}, \citet[]{genz11} have small uncertainties for both, and uncertainties were not listed for \citet{freun13}.

In order to make the weighting of each point reasonable (and to include what we believe to be more accurate estimates of the uncertainty), some adjustments were made to the data set.  First, the uncertainties on \citet{mied16} clump radius were scaled to be proportional to the average uncertainty of the \citet{wisn12} radii measurements since both use Keck/OSIRIS at similar redshift.  Second, 10$\%$ error\footnote{The average uncertainty for the unlensed data is $\sim15\%$. $10\%$ was used for these measurements so as to not underweight data points which may have lower uncertainties than the average based on methods or redshift.} was added to both the clump radius and luminosity of the $z\approx0$ \hii regions as well as the data from \citet{genz11} and \citet{freun13} to make the weighting of these data points consistent with surveys of similar objects in the PyStan fit. Lastly, the \citet{ngu16} SFRs were measured using 21 cm continuum emission and CO 1-0 data with an assumed typical uncertainty of 50$\%$ (with variation from 30$\%$ to 100$\%$) on the full sample of GMCs, MCCs, and galaxies used in their study. Since we are only using the nearby MCCs observed by \citet{ngu16} we apply an uncertainty of 40$\%$ error for these clump radius and luminosity measurements. It should be noted that the measured uncertainties do not account for differences in methods of detecting clumps and defining their sizes. This is a significant source of additional uncertainty discussed in detail in \citet{liv12, wisn12}\footnote{\citet{wisn12} estimates an additional 30\% uncertainty on clump sizes due to the method of measuring the clump size as well as resolution and systematic effects. We do not include this in our fitting as it would be the same additional weighting for all points and therefore not impact the overall fitting.}.

After these adjustments to the reported uncertainties were made the PyStan fitting was performed using a simple linear model:

\begin{equation}
ln(L_{H\alpha}) = \alpha \,ln(r_{clump}) + \beta
\end{equation}
where $\alpha$ and $\beta$ are the variables determined in the fit. Using this linear model required taking the natural logarithm of the data to produce a power-law fit of the form:

\begin{equation}
L_{H\alpha} = \exp(\beta) r_{clump}^{\alpha}
\end{equation}

The Stan multinormal function was used to fit this model with uncertainties on both \lha \, and $r_{clump}$, as well as allowing for intrinsic scatter in both dimensions. The multinormal function is a Hamiltonian Monte Carlo (HMC) method--a type of MCMC method which samples the derivatives of the probability density function \citep{stan17}. The geometry of the HMC is described further in \citet{bet11}.

The likelihood function used for a single data point in this model is:

\begin{equation}
    \mathcal{P}(\vec{x_{i}} | \mathcal{M}) = \int d(x_{th, i}) \mathcal{N}(\vec{x_{th,i}}, \Sigma + V | \vec{x}_i) \label{eqn:likelihood_i}\\
\end{equation}

With $\vec{x_{i}}=\{x_i, y_i\}, and \ \vec{x_{th}}=\{x_{th},y_{th}\}$, where $\vec{x_{th}}$ is the theoretical true positions of x and y ($x=r_{clump}, \ y=L_{H\alpha}$). $\mathcal{M}$ is the set of model parameters (slope, $\alpha$; intercept, $\beta$; and intrinsic scatter, $\sigma_x, \ \sigma_y$; prior values listed in Table \ref{tbl:priors}), $\Sigma$ corresponds to the covariance matrix with uncertainties on clump size and luminosity, and $V$  is a 2$\times$2 matrix incorporating intrinsic scatter\footnote{Note that the intrinsic scatter priors, $\sigma_x$ and $\sigma_y$ are squared in the matrix and therefore the resulting scatter values are absolute values and the distribution should be thought of as mirrored about zero.} ($V_{xx}=\sigma_x^2$, $V_{xy}=V_{yx}=\sigma_x \sigma_y$, $V_{yy}=\sigma_y^2$). $\mathcal{N}(\vec{x_{th}}, \Sigma + V | \vec{x}_i)$ is defined to be:
\begin{eqnarray}
& \mathcal{N}(\vec{x_{th}}, \Sigma + V | \vec{x}_i) \equiv \\ 
& \frac{1}{\sqrt{2 \pi \left|\Sigma + V\right|}} \exp{\left[-\frac{1}{2}(\vec{x}_i - \vec{x_{th}}) \cdot (\Sigma + V)^{-1} \cdot (\vec{x}_i - \vec{x_{th}}) \right]} \nonumber
\end{eqnarray}

The full likelihood function is found by summing Equation \ref{eqn:likelihood_i} over all data points:
\begin{equation}
    \mathcal{P}(\vec{x} | \mathcal{M}) = \prod_{i = 0}^{N-1} \mathcal{P}(\vec{x_{i}} | \mathcal{M})
\end{equation}

This model was also extended to three dimensions to investigate the dependence of the scaling relations on additional measured properties of the clumps. This gives a multi-parameter power-law fit of the form:
\begin{equation}\label{eqn:3D}
L_{H\alpha} = \exp(\beta) r_{clump}^{\alpha} \delta^{\gamma}
\end{equation}
with $\alpha$, $\beta$, and $\gamma$ being determined in the PyStan fitting and $\delta$ being an additional property of the clump such as velocity dispersion ($\sigma$) or host galaxy gas fraction (\fgas).
This is fit with the Stan multinormal function with uncertainties provided and intrinsic scatter allowed in all three dimensions.

\begin{table} 
\begin{center}
\caption{Priors used in PyStan Fitting}
\begin{tabular}{lcc}
\tableline
\tableline
Model Parameter & Minimum & Maximum \\ 
\tableline
slope, $\alpha$ & 0 & 5 \\
intercept, $\beta$ & 0 & 100 \\
scatter(r), $\sigma_x$ & 0 & 100 \\
scatter(L), $\sigma_y$ & 0 & 100 \\
*second slope, $\gamma$ & 0 & 5 \\
*scatter($\delta$), $\sigma_z$ & 0 & 100 \\
\tableline
\tableline
\label{tbl:priors}
\end{tabular}
\tablecomments{All priors used covered a significantly wider range than the values settled on after the warm-up phase of the fitting (those used in determination of model results), except for the scatter parameters which settle around a value of zero. However, these should be thought of as an absolute value mirrored about zero.\\
*: parameters used in extension of model to 3-D fits only.}
\end{center}
\end{table}

Note that the luminosity of the clumps in \ha \, (\lha) is used to investigate the star forming relations of the clumps since it is proportional to the star-formation rate (SFR) \citep{kenn98} and avoids differences in choice of initial mass function (IMF) between studies. Both \lha \, and  SFR are used to investigate clump scaling relations throughout the literature. When \sfrd \, is used in our analysis a Chabrier IMF \citep{chab03} is applied to convert from \lha \, for all data.

It should also be noted that the data set from \citet{mied16} consists of both resolved and unresolved\footnote{The unresolved clumps in \citet{mied16} give an upper limit on the size of these regions. These clumps have a 30\% uncertainty on their size included for weighting the data points and make up less than 2\% of the total sample. Therefore we do not expect an overestimate on the size of the clumps to have a significant impact on the resulting fits.} clumps. These will be denoted with different symbols in all plots but will be treated the same in the fitting. All $z\approx0$ \hii regions used in this paper \citep{kenn03, gal83, ars88, bas06, roz06, mon07} will be grouped together for the purposes of fitting and figures since they are all unlensed galaxies at $z\sim 0$ and have had corrections applied by \citet{wisn12}. The other local analogs \citep{fish16, ngu16} are grouped individually due to typically larger clump sizes and higher star formation rate densities (\sfrd) than the group of local \hii regions.

\section{Results: Clump Size and Star Formation Scaling Relations}\label{sec:results}

All data described in Section \ref{sec:data} and Table \ref{tbl:data_sets} were combined and divided into various subsets for fitting and investigating the clump size-luminosity relationship. This allows for the investigation of whether there is a dependence on redshift, study selection effects, velocity dispersion ($\sigma$) of the ionized gas in the clumps, star formation rate surface density (\sfrd), or gas fraction (\fgas) of the host galaxy. The results for each data subset are shown in Tables \ref{tbl:fit_params2D} - \ref{tbl:fit_params3DII}. These include the determined intercept, slope(s) and intrinsic scatter in each dimension as well as uncertainites on each of those values. The results of each fit discussed in the text as well as fits to additional data subsets (described in column 1) are included in these tables.

\begin{table*}[h!] 
\begin{center}
\caption{Size - Luminosity Relation Fit Parameters: ($L_{H\alpha} = e^{\beta} r_{clump}^{\alpha}$)}
\begin{tabular}{lrrrrrrr}
\tableline
\tableline
Data Set & Reference \#'s* & Figure & $\alpha$ & $\beta$ & Scatter (r) & Scatter (L) & \# of Clumps\\
\tableline
all data & 1-12 & \ref{fig:all_fit} & $3.029^{+0.027}_{-0.027}$ & $74.384^{+0.122}_{-0.126}$ & $0.186^{+0.124}_{-0.128}$ & $0.194^{+0.125}_{-0.127}$ & 2848\\
no $z\approx0$ \hii regions & 1-9, 11-12 & \nodata & $1.959^{+0.040}_{-0.037}$ & $82.644^{+0.257}_{-0.255}$ & $1.115^{+0.877}_{-0.822}$  & $1.246^{+0.778}_{-0.912}$ & 356\\   
high \sfrd \, (all z) & 1-12 & \ref{fig:SFRD_break} & $1.741^{+0.060}_{-0.067}$ & $85.159^{+0.377}_{-0.321}$ & $0.476^{+0.355}_{-0.333}$ & $0.484^{+0.354}_{-0.324}$ & 152\\
low \sfrd \, (all z) & 1-12 & \ref{fig:SFRD_break} & $2.767^{+0.021}_{-0.023}$ & $75.356^{+0.100}_{-0.104}$ & $0.121^{+0.086}_{-0.073}$ & $0.136^{+0.076}_{-0.086}$ & 2696\\
high \sfrd \, (z$\sim$0) & 10-12 & \nodata & $1.479^{+0.094}_{-0.052}$ & $86.416^{+0.260}_{-0.504}$ & $0.940^{+0.769}_{-0.629}$ & $1.021^{+0.916}_{-0.666}$ & 114\\
low \sfrd \, (z$\sim$0) & 10-12 & \nodata & $2.656^{+0.034}_{-0.034}$ & $75.798^{+0.149}_{-0.153}$ & $0.138^{+0.097}_{-0.095}$ & $0.143^{+0.091}_{-0.097}$ & 2527\\
corrected high \sfrd & 1-12 & \ref{fig:corrected} & $1.725^{+0.067}_{-0.059}$ & $85.334^{+0.327}_{-0.364}$ & $0.502^{+0.453}_{-0.344}$ & $0.607^{+0.359}_{-0.411}$ & 200 \\
corrected low \sfrd & 1-12 & \ref{fig:corrected} & $2.862^{+0.034}_{-0.037}$ & $74.953^{+0.165}_{-0.156}$ & $0.122^{+0.080}_{-0.081}$ & $0.121^{+0.081}_{-0.080}$ & 2648\\
corrected; no $z\approx0$ \hii regions & 1-9, 11-12 & \nodata & $2.296^{+0.070}_{-0.077}$ & $81.230^{+0.386}_{-0.396}$ & $0.639^{+0.488}_{-0.406}$ & $0.636^{+0.460}_{-0.440}$ & 356\\
$z\approx0$ \hii regions only & 10 & \ref{fig:redshift} & $2.448^{+0.036}_{-0.034}$ & $76.681^{+0.157}_{-0.160}$ & $0.198^{+0.123}_{-0.131}$ & $0.179^{+0.134}_{-0.123}$ & 2492\\ 
all $z \sim 0$ & 10-12 & \ref{fig:redshift} & $3.057^{+0.038}_{-0.035}$ & $74.229^{+0.148}_{-0.165}$ & $0.176^{+0.119}_{-0.121}$ & $0.174^{+0.120}_{-0.120}$& 2641\\
$0.6 \leq z < 1.5 $ & 3-5, 7-9 & \ref{fig:redshift} & $2.099^{+0.078}_{-0.068}$ & $80.498^{+0.457}_{-0.519}$ & $0.318^{+0.221}_{-0.203}$ & $0.328^{+0.227}_{-0.217}$ & 160\\
$z\geq 1.5$ & 1-4, 6 & \ref{fig:redshift} & $1.828^{+0.180}_{-0.080}$ & $84.175^{+0.626}_{-1.281}$ & $1.959^{+1.828}_{-1.334}$ & $2.020^{+1.852}_{-1.452}$ & 47\\
lensed high-z & 1-5 & \ref{fig:lensed} & $2.099^{+0.199}_{-0.147}$ & $81.188^{+0.859}_{-1.230}$ & $0.790^{+0.750}_{-0.566}$ & $0.804^{+0.706}_{-0.548}$ & 108\\
unlensed high-z & 6-9 & \ref{fig:lensed} & $2.266^{+0.115}_{-0.086}$ & $79.465^{+0.756}_{-0.867}$ & $0.414^{+0.397}_{-0.285}$ & $0.488^{+0.324}_{-0.326}$ & 209\\
\tableline
\label{tbl:fit_params2D}
\end{tabular}
\tablecomments{*: Reference numbers correspond to data from studies as defined in Table \ref{tbl:data_sets}.}
\end{center}
\end{table*}

\begin{table*}[h!]
\begin{center}
\caption{3D Fit Parameters: $\sigma$ ($L_{H\alpha} = e^{\beta} r_{clump}^{\alpha} \sigma_{clump}^{\gamma}$)}
\begin{tabular}{lrrrrrrr}
\tableline
\tableline
Data Set (Reference \#'s) & $\alpha$ & $\gamma$ & $\beta$ & Scatter (r) & Scatter ($\sigma$) & Scatter (L) & \# of Clumps\\
\tableline
2,3,6,7,9,10*,11 & $1.026^{+0.089}_{-0.086}$ & $2.211^{+0.141}_{-0.138}$ & $79.038^{+0.377}_{-0.492}$ & $0.091^{+0.094}_{-0.062}$ & $0.091^{+0.094}_{-0.062}$ & $0.098^{+0.095}_{-0.067}$ & 346\\
2,3,6,7,9,10*,11 (2D) & $2.049^{+0.044}_{-0.036}$ & \nodata & $81.531^{+0.240}_{-0.302}$ & $1.539^{+0.936}_{-0.992}$ & \nodata & $1.246^{+1.033}_{-0.918}$ & 346\\
\tableline
\label{tbl:fit_params3DI}
\end{tabular}
\tablecomments{*:Only \protect\citet{gal83, ars88, bas06, roz06, mon07} from this set number. Not all data sets in the full sample included measurements of $\sigma_{clump}$, leading to slightly higher uncertainties on the fit. The results of fitting this sample with the 3D model above are in the first row and the 2D fit excluding $\sigma$ are in the second row for comparison of the change in slope, uncertainty, and scatter when including this third dimension in the fit.}
\end{center}
\end{table*}

\begin{table*}[h!]
\begin{center}
\caption{3D Fit Parameters: \fgas ($L_{H\alpha} = e^{\beta} r_{clump}^{\alpha} f_{gas}^{\gamma}$)}
\begin{tabular}{lrrrrrrr}
\tableline
\tableline
Data Set (Reference \#'s) & $\alpha$ & $\gamma$ & $\beta$ & Scatter (r) & Scatter (\fgas) & Scatter (L) & \# of Clumps\\
\tableline
5,8,9,11* & $1.345^{+0.087}_{-0.092}$ & $0.471^{+0.064}_{-0.064}$ & $86.716^{+0.666}_{-0.629}$ & $0.412^{+0.455}_{-0.298}$ & $0.412^{+0.455}_{-0.298}$ & $0.370^{+0.477}_{-0.277}$ & 157\\ 
5,8,9,11* (2D) & $1.611^{+0.030}_{-0.030}$ & \nodata & $84.942^{+0.289}_{-0.269}$ & $2.223^{+1.555}_{-1.505}$ & \nodata & $2.299^{+1.469}_{-1.502}$ & 157\\
\tableline
\label{tbl:fit_params3DII}
\end{tabular}
\tablecomments{*: Measurements of \fgas \, from \protect\citet{white17}, size and luminosity from \protect\citet{fish16}. Not all data sets included measurements of \fgas, leading to slightly higher uncertainties on the fit. The results of fitting this sample with the 3D model above are in the first row and the 2D fit excluding \fgas \, are in the second row for comparison of the change in slope, uncertainty, and scatter when including this third dimension in the fit.}
\end{center}
\end{table*}

\begin{figure*}[h]
\epsscale{.92}
\plotone{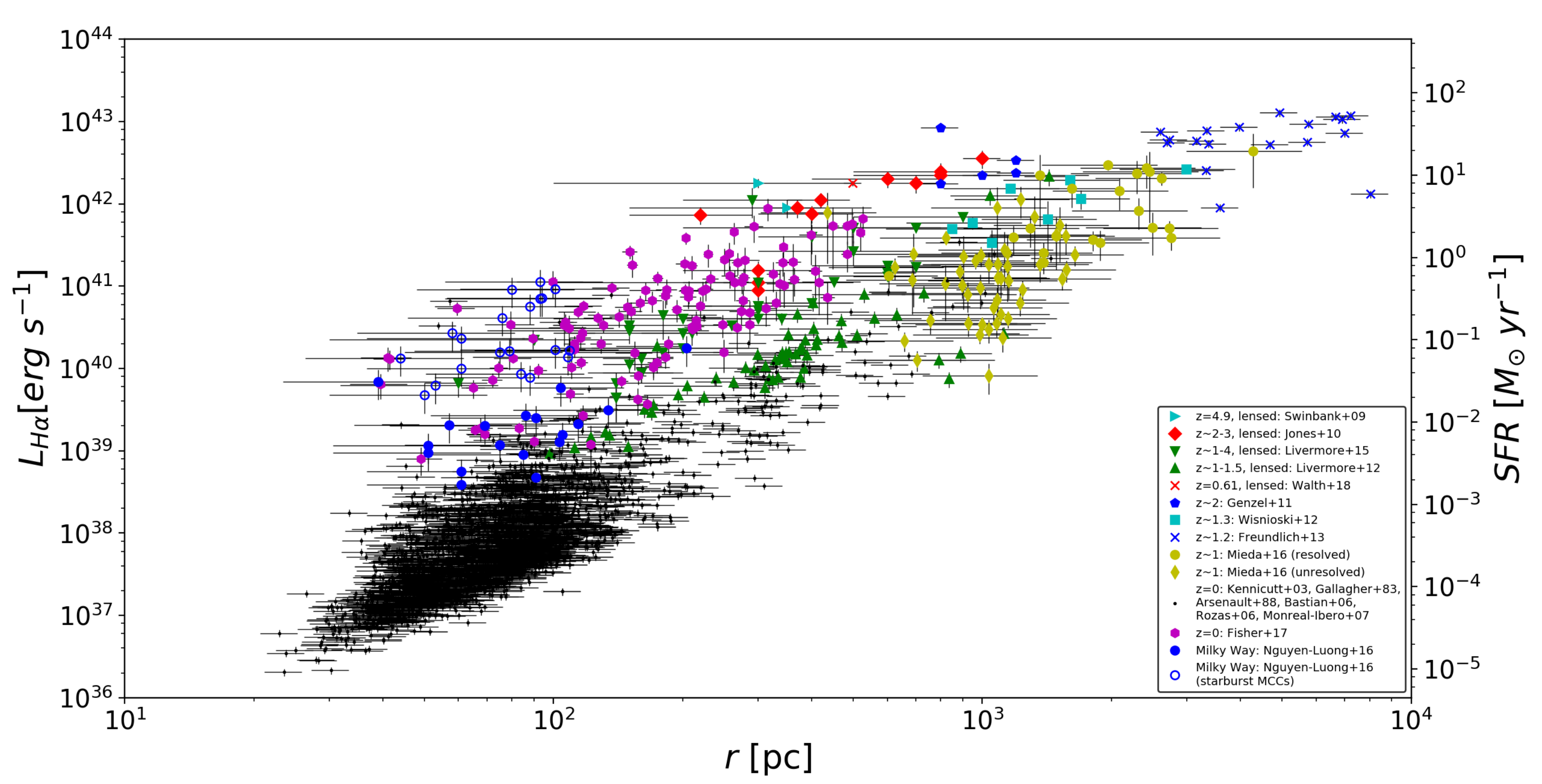}
\caption{Clump size and luminosity for all the data used throughout this paper. In the case that the star formation rate (SFR) only is reported, this is converted back to the equivalent \ha \, luminosity following \protect\citet{kenn98} and the initial mass function from \protect\citet{chab03}. This was the case for the \protect\citet{swin09, liv12, freun13, wal18} data. NOTE: The SFR reported in \protect\citet{swin09, freun13} is derived from [\oii] emissions, not $H\alpha$, which may introduce up to a factor of $\sim$2 difference from \ha \, derived SFR \protect\citep{kew04}. The size reported for \protect\citet{freun13} clumps is derived from IRAM CO measurements and is sometimes less than the 1" slit used for [\oii] luminosity measurements. \protect\citet{ngu16} use CO 1-0 and 21 cm continuum emission to estimate SFR which can contribute to the scatter between these measurements and those from ionized gas emission. However, the 40\% uncertainty for these data points significantly reduces their weight in the fit. \label{fig:all_fit}}
\end{figure*}

\begin{figure*}[h]
\epsscale{0.92}
\plotone{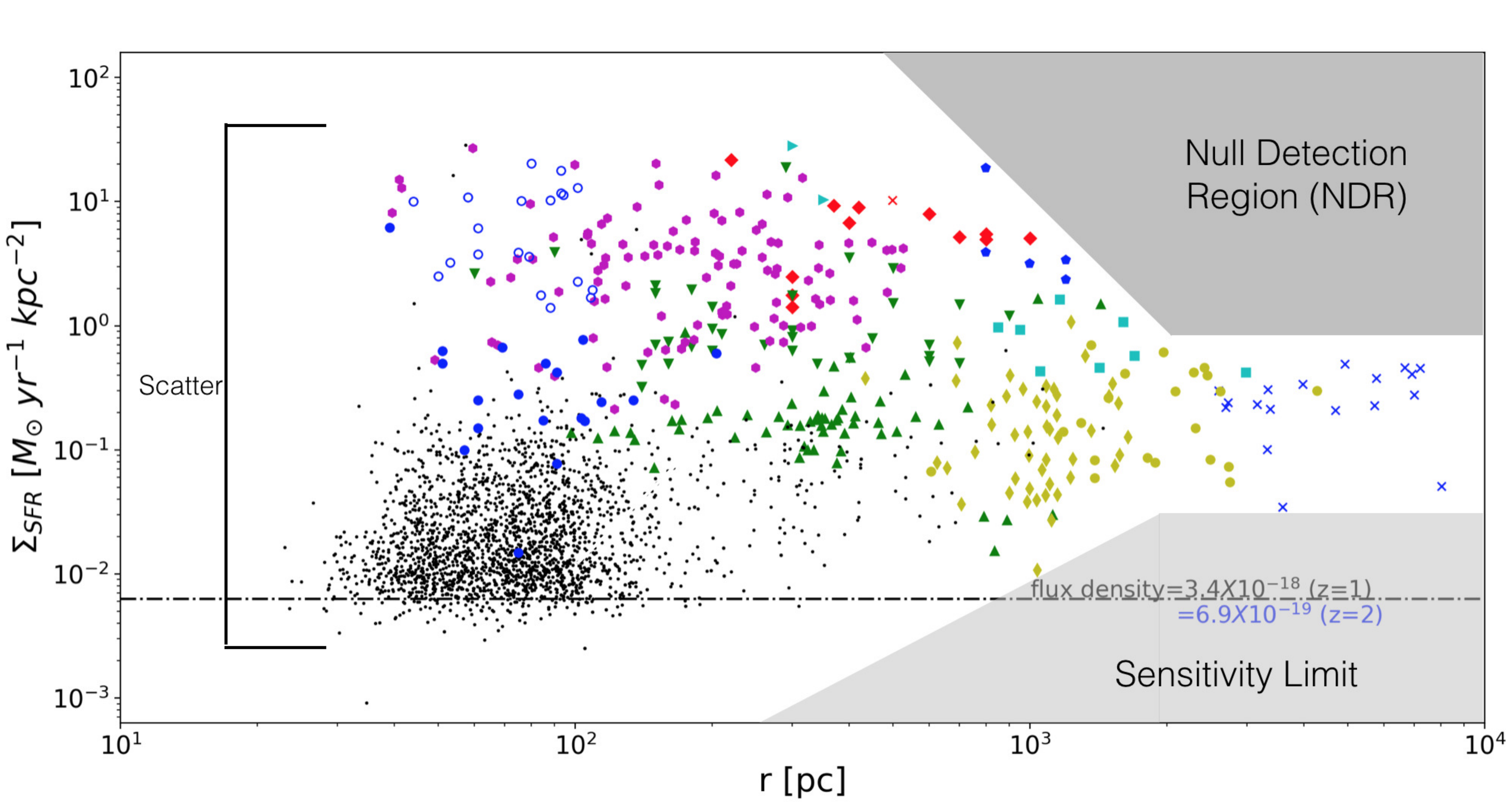}
\caption{Clump size plotted against the star formation surface density ($\Sigma_{SFR}$) to illustrate the large variation in the various data sets (see Figure \ref{fig:all_fit} for legend). The scatter is outlined to the left; this is prevalent for both the Milky Way up to high-redshift. This indicates that the clump size is not the only factor influencing the SFR. The shaded region in the lower right illustrates the lack of data seen at this regime of large, low surface brightness clumps which is likely due to a sensitivity limit of the instruments being used.  The dashed black line shows what the observed flux density would be at this \sfrd \, for $z=1$ (black text) and $z=2$ (blue text). This exact limit will depend on the individual study and vary within studies in the case of gravitationally lensed galaxies (see Figure \ref{fig:sensitivity} for more detailed sensitivity levels). The shaded region in the upper right labelled ``Null Detection Region (NDR)" corresponds to a lack of observations of large clumps with \textit{high} surface brightness. This would not be due to a sensitivity limit and likely corresponds to a physical absence of clumps in this regime. \label{fig:SFRDcartoon}}
\end{figure*}

The overall combined data set shown in Figures \ref{fig:all_fit} and \ref{fig:SFRDcartoon} results in a scaling relationship of \lha $\propto r^{3.029}$. This sample includes a wider range of data than has previously been used in this type of comparison with these figures illustrating some key features of the data set. The large scatter shown in the size-luminosity plot of Figure \ref{fig:all_fit} and highlighted in Figure \ref{fig:SFRDcartoon} causes one of the main problems with determining a reliable size-luminosity relationship. Different relationships will be derived depending on what data is used for the comparison, which could account for some of the variation seen in previous studies. The large scatter ($\sim$3dex) at fixed radius illustrated in Figure \ref{fig:SFRDcartoon} indicates dependence of the luminosity on a second parameter in addition to the radius of the clump. In order to investigate the reasons for this scatter and what drives the relationship we have divided the data into the subsets shown in Table \ref{tbl:fit_params2D} and described in the following pages. 

The absence of data in the lower right of Figures \ref{fig:SFRDcartoon} \& \ref{fig:all_fit} (corresponding to large, low surface brightness clumps) is likely due to a sensitivity limit in what clumps can be observed with current instruments. This is discussed further in Section \ref{sec:bias} and may be partially responsible for the steeper slope here than determined in previous studies. In contrast to this, the lack of observed data with large, high surface brightness clumps cannot be due to a sensitivity limit. This corresponds to the shaded region in the upper right of Figure \ref{fig:SFRDcartoon} referred to as the ``Null Detection Region". This may be due to a physical absence of clumps at this regime which could be the result of feedback mechanisms (discussed further in Section \ref{sec:feedback}).

\subsection{Star Formation Surface Density (\sfrd) Break} \label{sec:SFRDbreak}

\citet{ngu16} determine that there is a break in the slope of the scaling relations and star formation laws locally in their sample of MCCs between normal star-forming objects and what they refer to as mini-starbursts (gravitationally unbound MCC's with \sfrd \, $\rm >1 \ M_{\sun} \ yr^{-1} \ kpc^{-2}$). \citet{john17a} find that \hii regions in the SINGS sample \citep{kenn03} have significantly lower \sfrd \, than the z$\sim$2 lensed samples they are comparing them to and that the higher \sfrd \, of the DYNAMO galaxies \citep{fish16} provide a better analog to the massive star forming clumps seen at high redshift. This indicates that there may be two different process occurring in different types of clumps with different scaling relations that skew the results of fitting the data as a whole.

\begin{figure*}[h]
\epsscale{2.0}
\plottwo{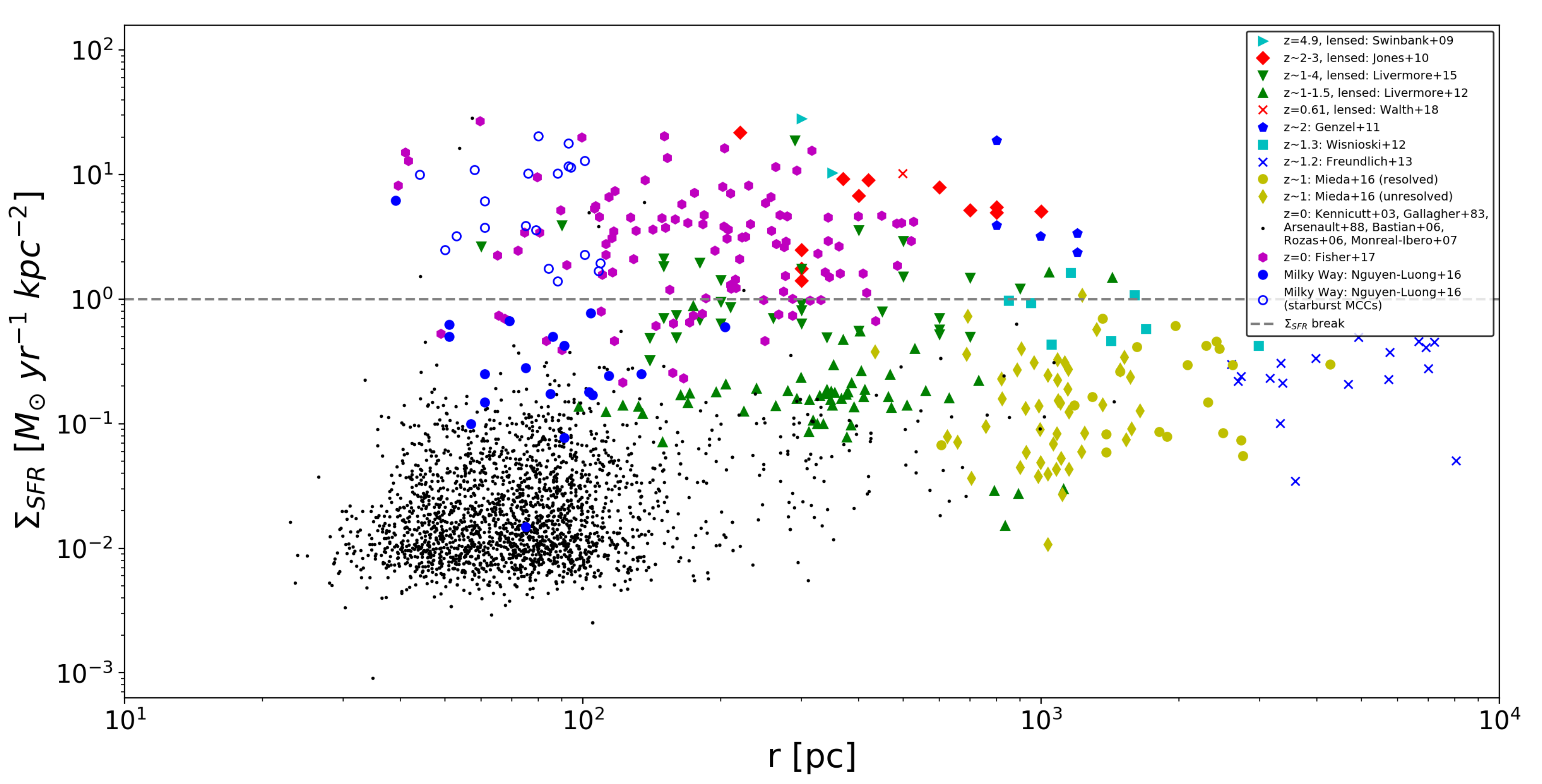}{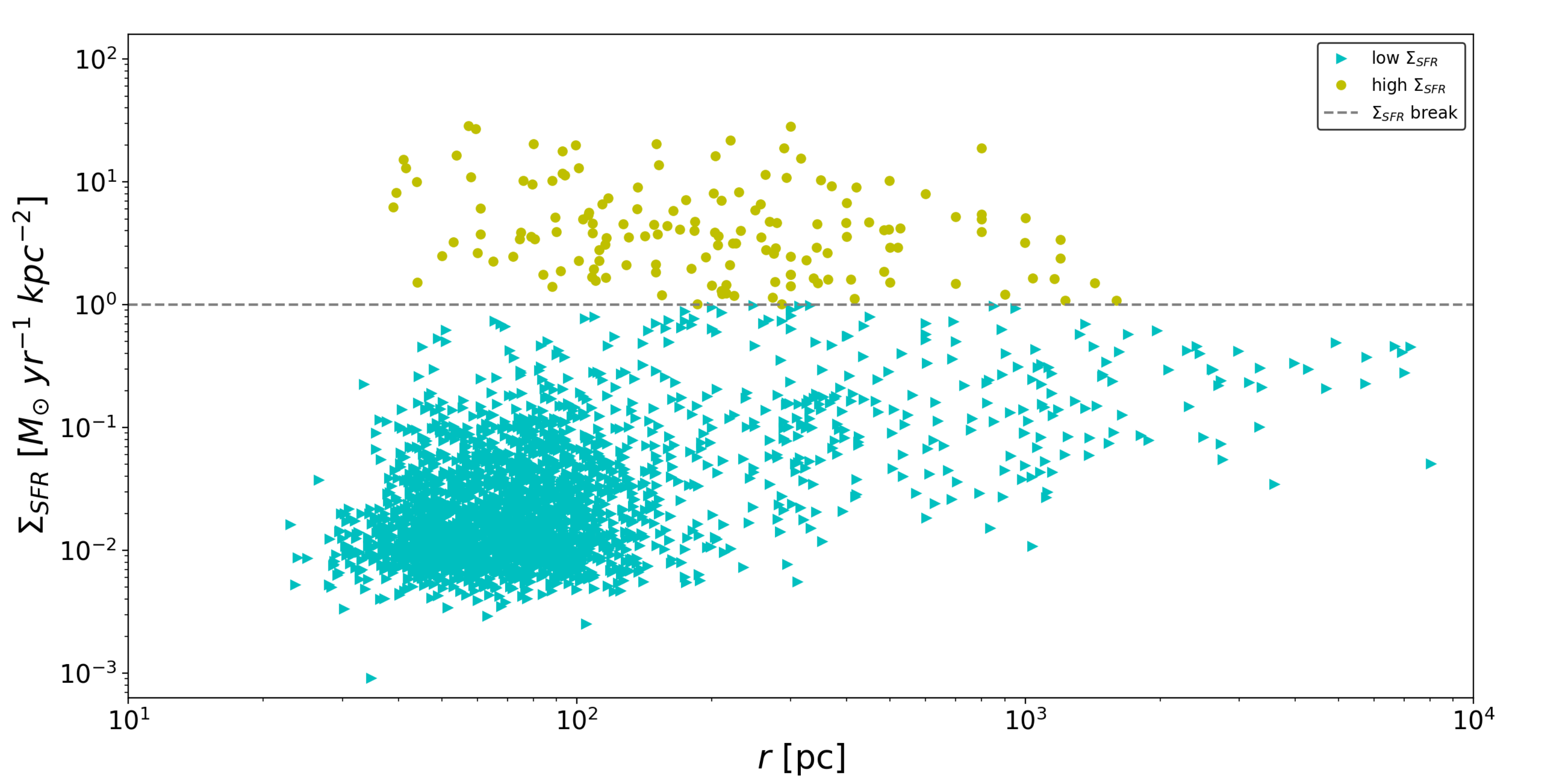}
\caption{Clump size plotted against \sfrd \, illustrating the power-law break at \sfrd \, $\rm = 1 \, M_{\sun} \ yr^{-1} \ kpc^{-2}$ (grey dashed line). The top figure shows the data separated by study, while the bottom figure is divided into low and high \sfrd. \label{fig:SFRD}}
\end{figure*}

In order to test this data were divided into two groups, high \sfrd \, and low \sfrd \, defined by varying \sfrd \, cut-offs. We investigate the location of the break by incrementally varying the cut-off \sfrd \, value and comparing the slope for the high and low \sfrd \, subsets to a baseline result with the break defined at \sfrd \,$\rm = 1 \ M_{\sun} \ yr^{-1} \ kpc^{-2}$ as illustrated in Figure \ref{fig:SFRD}. We vary the \sfrd \, break in increments of 0.25 between \sfrd \, $\rm = 0.25 - 2.5 \ M_{\sun} \ yr^{-1} \ kpc^{-2}$, and then adjust the step size due to the logarithmic nature of the distribution\footnote{For example one step below \sfrd \, $\rm = 0.25 \ M_{\sun} \ yr^{-1} \ kpc^{-2}$ would shift to including the full sample in the fitting, adding a significant number of data points and scatter. To ensure any change in slope is due to a real change in location of the power law break and not the increase in data points we decrease the step size below \sfrd \, $\rm = 0.25 \ M_{\sun} \ yr^{-1} \ kpc^{-2}$.}. Changes in slope $>0.12$ (3$\times$ the average uncertainty in the baseline slope) are considered significant, but do not result from breaks between \sfrd \, $\rm = 0.25 \ M_{\sun} \ yr^{-1} \ kpc^{-2}$ and \sfrd \, $\rm = 1.25 \ M_{\sun} \ yr^{-1} \ kpc^{-2}$. With \sfrd \, breaks located outside of this range the resulting slopes deviate more rapidly and by more than 0.12 from the baseline, supporting the break location in this \sfrd \, phase space. Further data at large clump sizes will help to constrain this break in the future. For simplicity we discuss the fitting results only with the break at \sfrd \, $\rm = 1 \ M_{\sun} \ yr^{-1} \ kpc^{-2}$. While this is only an approximate value for the cut-off, the resulting scaling relations for the high and low \sfrd \, bins are consistent with other cut-offs in this region. Dividing the full data set into high and low \sfrd \, clumps results in different slopes, which may imply two unique clump populations with different physical processes occurring:
\begin{eqnarray}
\begin{array}{cc}
\Sigma_{SFR} > 1 \ \rm M_{\sun} \ yr^{-1} \ kpc^{-2}: L_{H\alpha} \propto r_{clump}^{1.7} \nonumber\\
\Sigma_{SFR} \leq 1 \ \rm M_{\sun} \ yr^{-1} \ kpc^{-2}: L_{H\alpha} \propto r_{clump}^{2.8}
\end{array}
\end{eqnarray}

\begin{figure*}[h]
\epsscale{2.0}
\plottwo{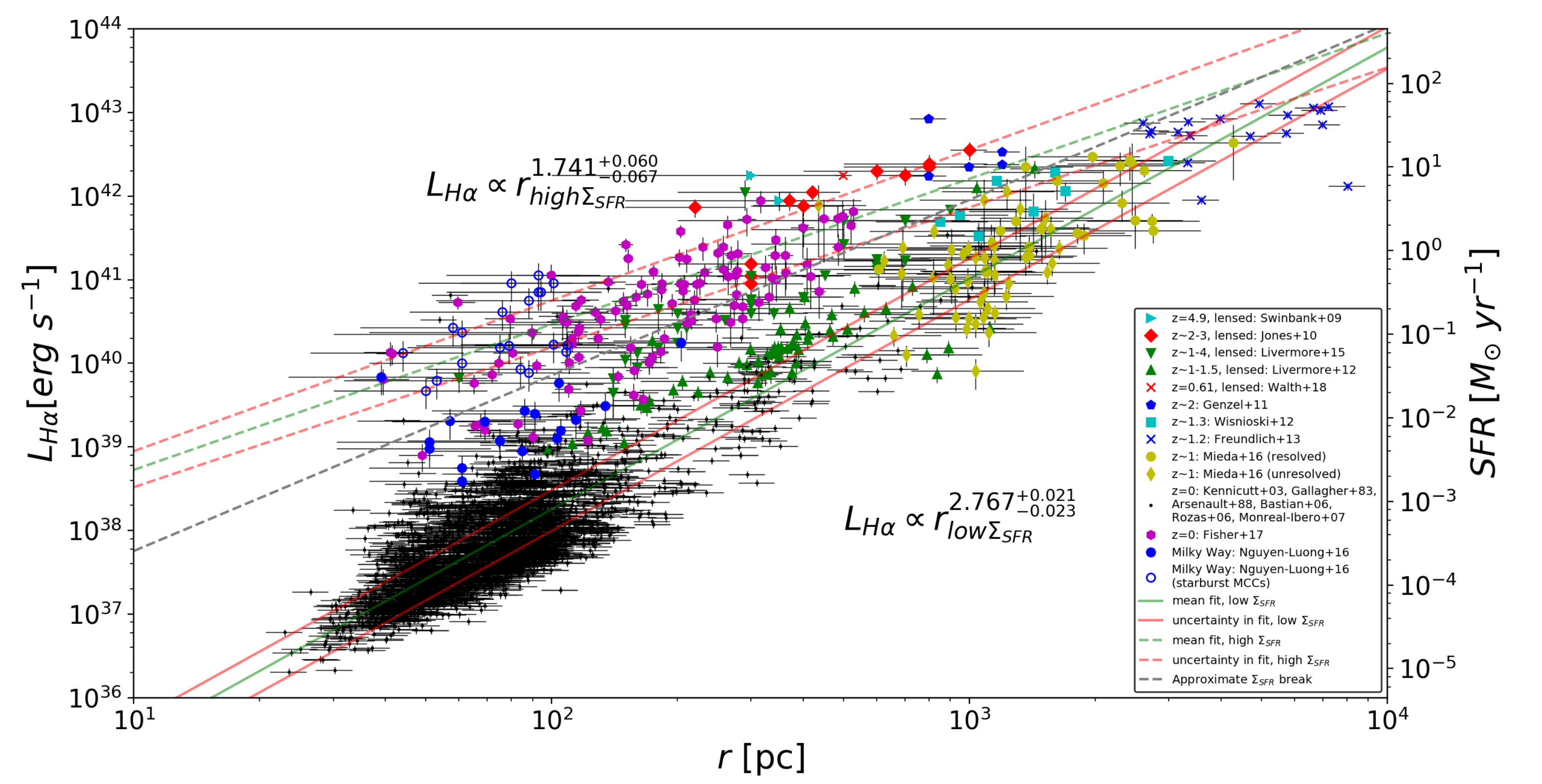}{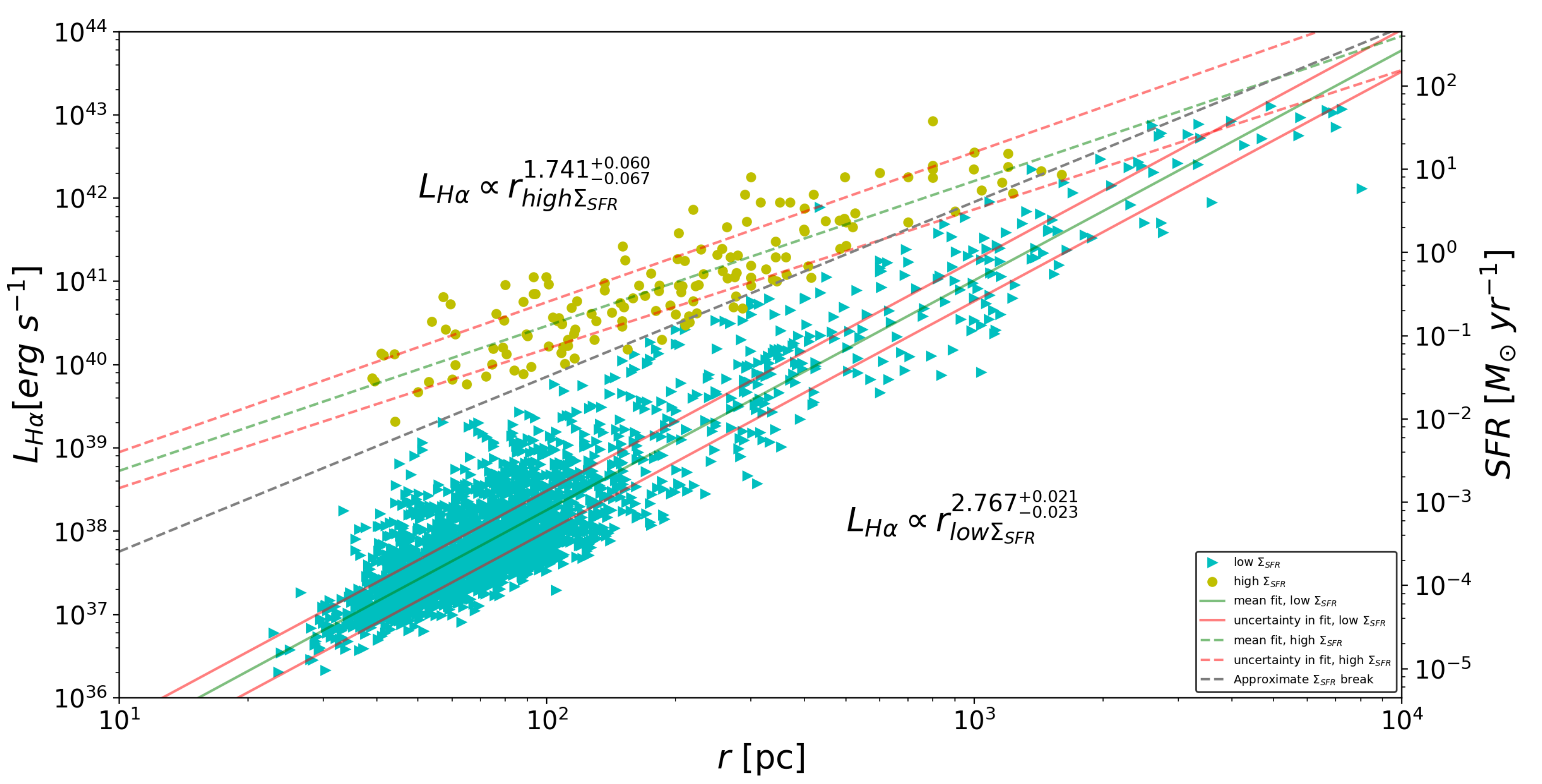}
\caption{Clump size and luminosity relation for the high and low  \sfrd \, bins. The dashed red and green lines show the best fit to the high \sfrd \, data, while the solid red and green lines show the best fit of the low \sfrd \, data. The grey dashed line is approximately where the \sfrd \, $\rm = 1 \ M_{\sun} \ yr^{-1} \ kpc^{-2}$ cut-off lies when converted to luminosity. The top figure shows the data separated by study, while the bottom figure is divided into low and high \sfrd.\label{fig:SFRD_break}}
\end{figure*}

The difference in these relationships and approximately where the cut-off lies on the size-luminosity plot are shown in Figure \ref{fig:SFRD_break}.  Interestingly, the higher \sfrd \, data scales with $r^{1.7}$ which is near to what has been suggested for clump formation driven by Toomre instability \citep[$L \propto r^2$ by extending the equations given in][]{genz11}, while the lower \sfrd \, data scales like  $r^{2.8}$, closer to the expected relation if the clumps are represented by Str\"{o}mgren spheres ($L \propto r^3$) \citep{wisn12}. However, the true slope may be shallower than what we find here if lower surface brightness clumps are not being detected due to sensitivity limits. As is shown in the bottom portion of Figure \ref{fig:SFRD_break} the division of the data into high and low \sfrd \, sets results in two separate regions on the size-luminosity plot with very little overlap due to scatter. This further supports the idea of multiple processes occurring in these two clump populations even with the possible sensitivity limit.

\begin{figure*}[h]
\epsscale{1.6}
\plottwo{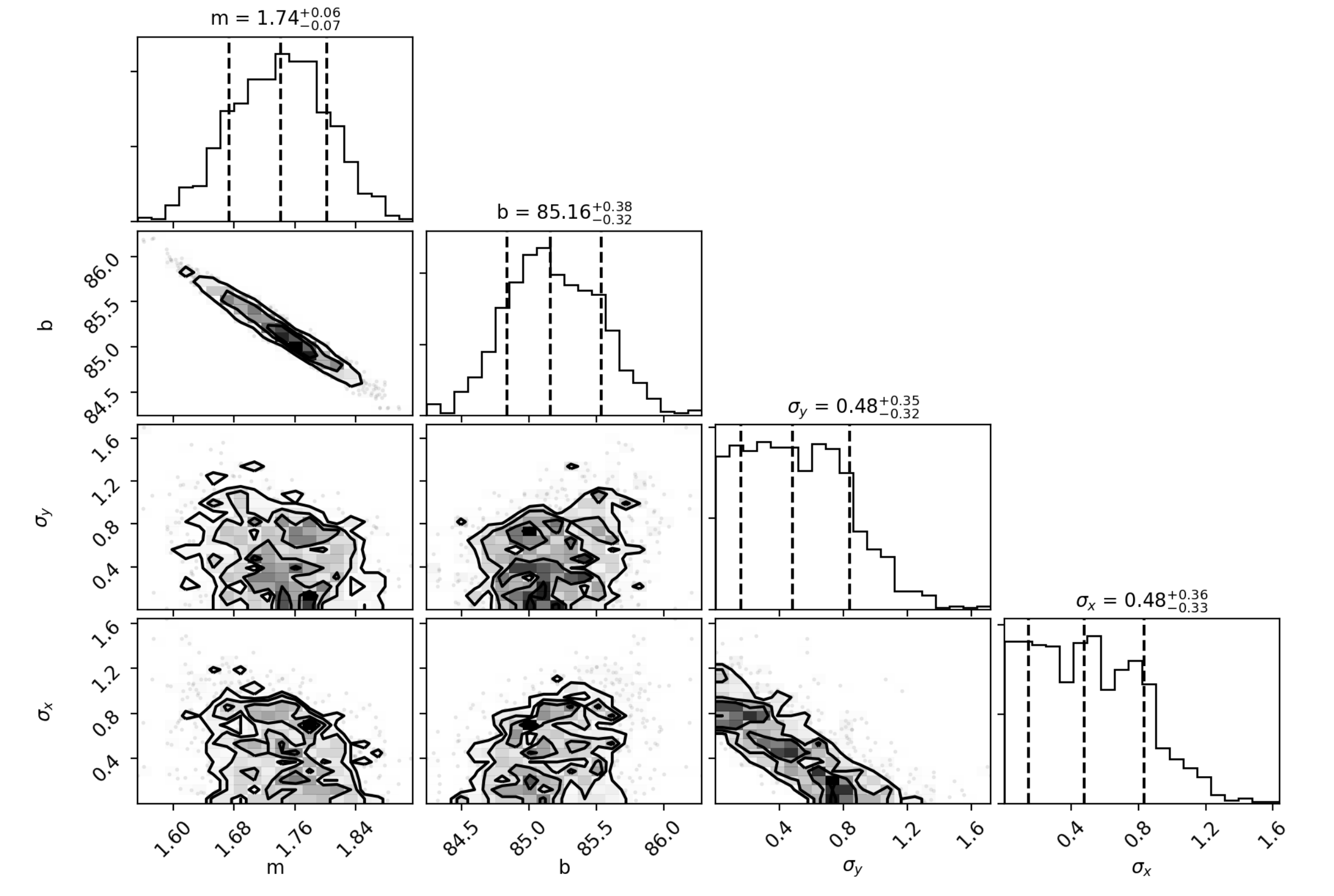}{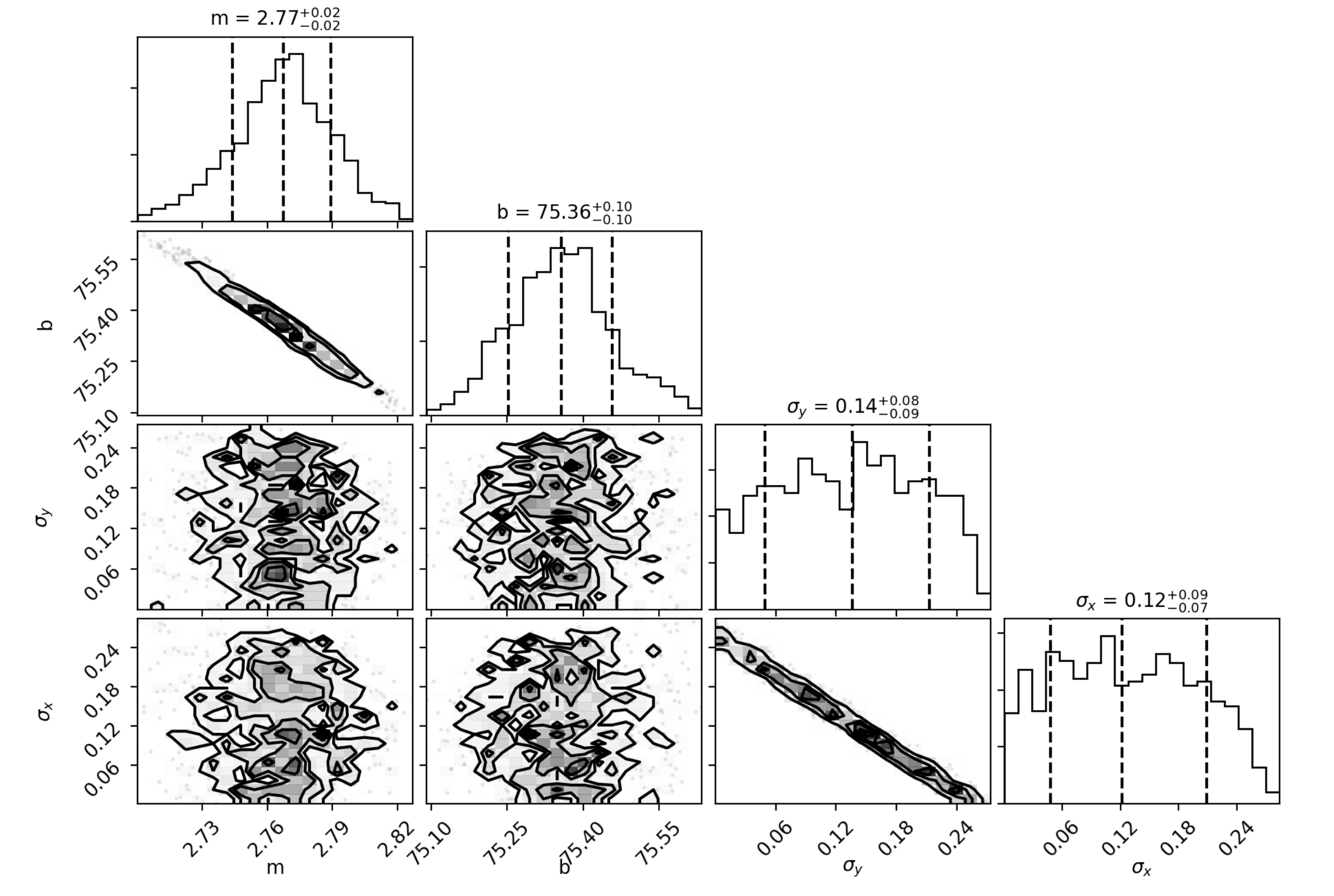}
\caption{Posterior probability distribution for the fits shown in Figure \ref{fig:SFRD_break} with a power-law break based on \sfrd. The top figure shows the distribution for the fit to the data with \sfrd \, $\rm > 1 \, M_{\sun} \ yr^{-1} \ kpc^{-2}$ while the bottom figure corresponds to the fit to the data with \sfrd \, $\rm < 1 \, M_{\sun} \ yr^{-1} \ kpc^{-2}$. $\sigma_x$ and $\sigma_y$ are intrinsic scatter parameters corresponding to \rha \, and \lha \, respectively. Both the slope and intercept of these fits are well constrained from a much broader range of priors (Table \ref{tbl:priors}). The values for intrinsic scatter, $\sigma_x$ and $\sigma_y$, are not limited on the high end, but do tend towards zero. As scatter is an absolute value negative values are not possible and the distribution can be thought of as mirrored about zero. These small values of intrinsic scatter indicate that the scatter seen in the data is not intrinsic scatter but may be due to uncertainties. This posterior probability distribution is representative of what is produced for all the fits performed in this analysis. \label{fig:SFRD_prob}}
\end{figure*}

The scaling found when fitting the full data set with this power-law break is nearly the same as is found when applying the same break to only the $z\approx0$ data at the smaller size end of the sample (maximum clump size of 1.4 kpc vs. 8 kpc for the full sample) \citep{kenn03, gal83, ars88, bas06, roz06, mon07, fish16, ngu16}. With only this local data, a scaling relationship of $L_{H\alpha} \propto r^{2.7}$ is found for the low \sfrd \, star forming regions and $L_{H\alpha} \propto r^{1.5}$ for the high \sfrd \, star forming regions. The uncertainty and scatter on these fits is shown in Table \ref{tbl:fit_params2D} while the posterior probability distribution for the fit to the high and low \sfrd \, subsets of the full sample is displayed in Figure \ref{fig:SFRD_prob}.

\subsection{Corrections for Beam Smearing}\label{sec:blurring}

It has been suggested that the lower spatial resolution (see Table \ref{tbl:data_sets}) of unlensed high-redshift samples could lead to incorrectly measured clump sizes and an effect of observing ``clumps within clumps" where what is actually a group of smaller clumps is observed as one large clump due to beam smearing \citep{fish16, cav17}. To investigate what affect this may have on the measured clump properties, \citet{fish16} degrade the images of their local galaxies to match the resolution of $z\sim1-2$ observations (from $\sim100$ pc to $\sim800$ pc spatial resolution). They find that this typically leads to about a factor of 5 increase in the observed SFR (proportional to \lha) and a $\sim6\times$ decrease in the observed \sfrd \, (translating to a $\sim 5.5\times$ increase in clump sizes).  This effect of resolution has also recently been investigated by \citet{cav17} in multiple gravitationally lensed images of the same galaxy. The images divide into two distinct sets: the ``cosmic snake" which consists of four elongated images of the galaxy, and what is referred to as the counterimage. \citet{cav17} report a resolution limit of $\sim 300$ pc in the counterimage, but can get down to a scale of $\sim 30$ pc in the cosmic snake. They find that the clumps observed in the counterimage are typically a factor of 2-3 larger than those observed in the cosmic snake.

In order to determine if these effects were occurring and could be currently observed in unlensed galaxies we chose one of the brightest galaxies in the IROCKS sample \citep{mied16} to re-observe at a smaller plate scale. The original observations made use of the 0.1" plate scale on the OSIRIS instrument at Keck in order to maximize the surface brightness sensitivity (hence, the choice of a high surface brightness galaxy).

\begin{figure*}[h]
\gridline{\fig{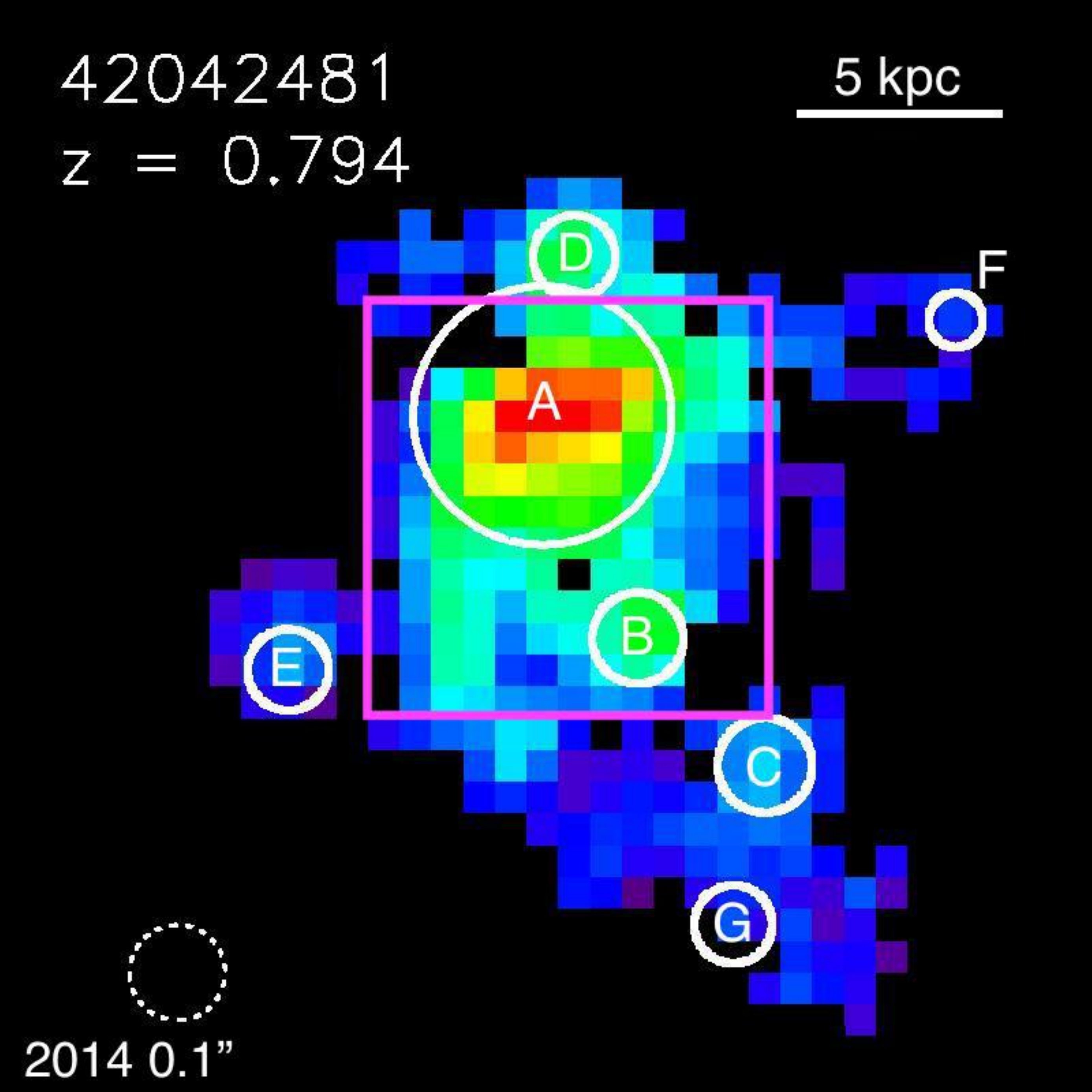}{0.3\textwidth}{(a): 2014 0.1" observations}
        \fig{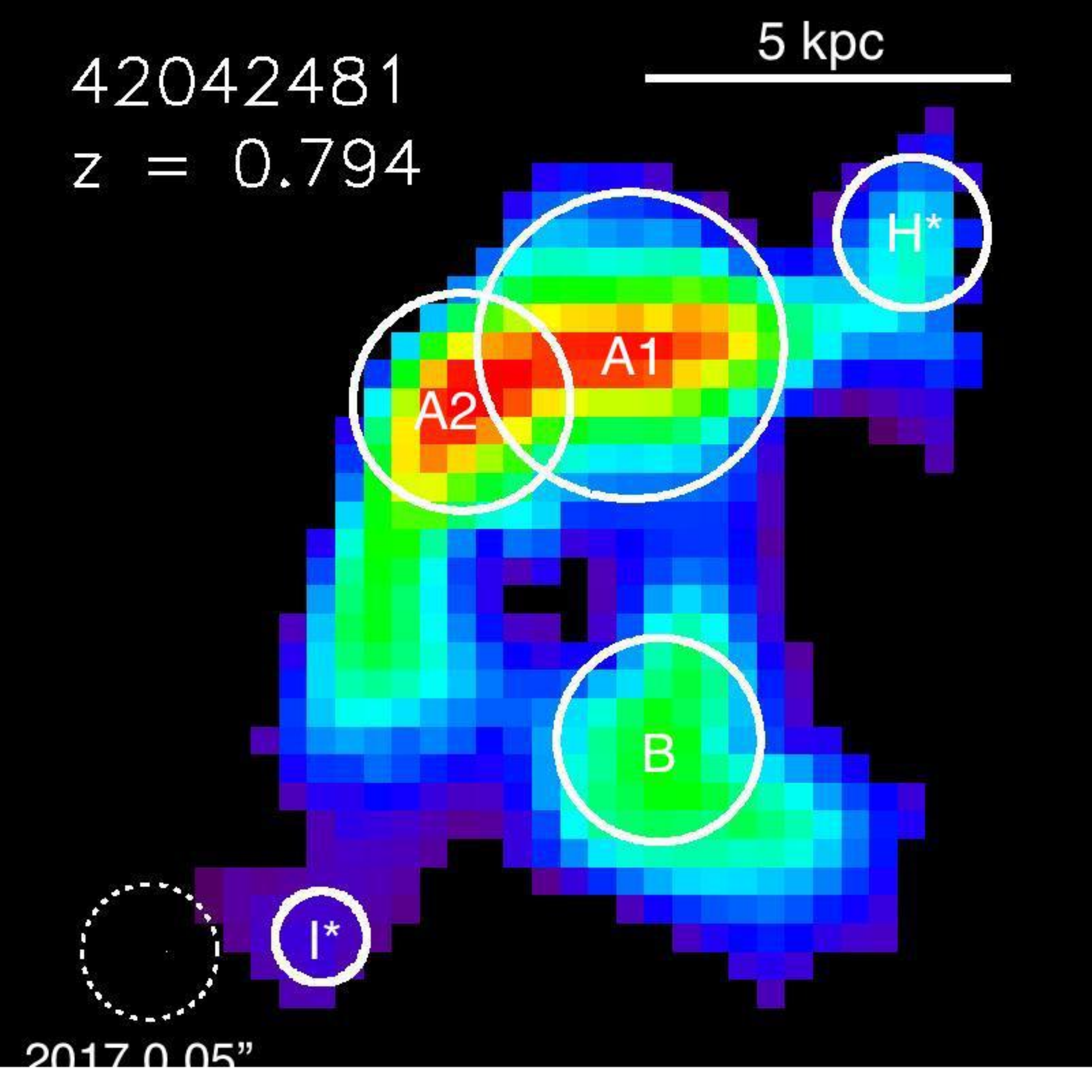}{0.3\textwidth}{(b): 2017 0.05" observations}
        \fig{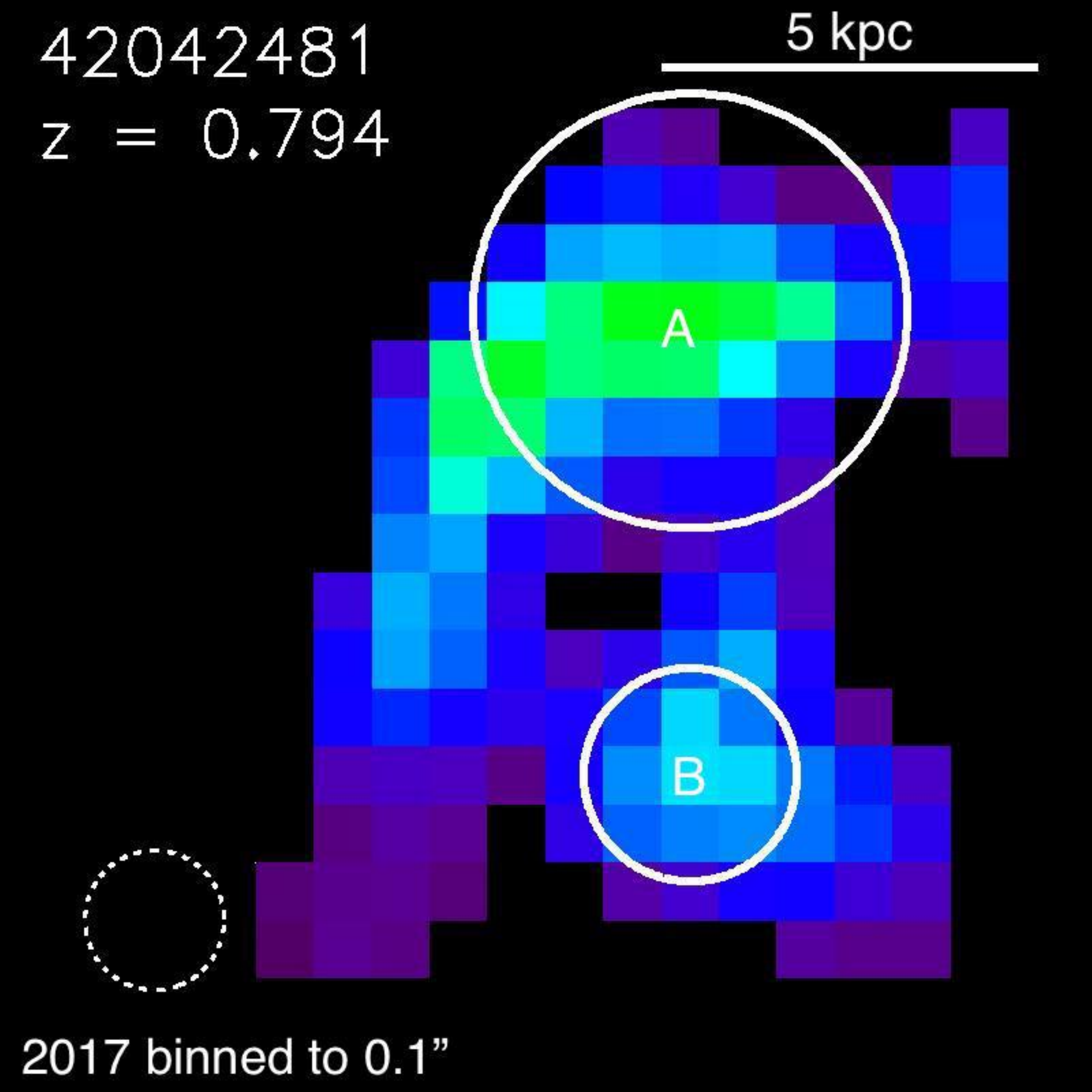}{0.3\textwidth}{(c): 2017 0.05" observations binned to 0.1"}}
\caption{Clump locations and sizes identified from \ha \, flux. Following the definition of \protect\citet{mied16}, clumps are located via a local \ha \, peak separated by at least two pixels from a neighboring peak. Clumps A and B are at the same location in all panels; clumps H* and I* in panel (b) are new clumps not found in the initial 0.1" observations. Clump A in the 0.1" plate scale observations appears to break up into two clumps at higher resolution. Sizes of all clumps are listed in Table \ref{tbl:clump_sizes}. A scale bar is located at the top right of each figure; note that the pixel scale is different for the 0.1" and 0.05" observations. The pink box in panel (a) shows the region covered by the 2017 0.05" plate scale observations. The dashed circle in the lower left of each panel shows the PSF size for that night of observations. \label{fig:Ha_maps}}
\end{figure*}

Object 42042481 was observed on 2017 August 12 with Keck/OSIRIS at a plate scale of 0.05" per spaxel and the narrowband J filter. Seven 900s exposures (giving 1.75h total integration time; as opposed to 2.5h total integration time at 0.1") of 42042481 were taken along with a pure sky frame. The data was reduced using the OSIRIS data reduction pipeline (DRP) version 4.1 producing a combined cube of all seven frames. This cube was also binned down to the spatial resolution of the 0.1" plate scale for an additional comparison along with the initial observations. These cubes were spatially smoothed in the manner described in \citet{mied16} and to an equivalent FWHM before the same custom IDL scripts were used to determine the locations and sizes of \ha \, clumps (this process was also repeated by MC on the previous observation of 42042481 to ensure a consistent comparison). The resulting \ha \, maps for the original 0.1" observations, the 0.05" observations, and the binned data are shown in Figure \ref{fig:Ha_maps} with marked clump locations and size of the point spread function (PSF). The properties of these clumps are reported in Table \ref{tbl:clump_sizes}.

The shift from the original 0.1" to 0.05" plate scale resulted in an improved spatial resolution limit from $\sim800$ pc to $\sim400$ pc causing the largest clump to split into two clumps each roughly half the size originally measured. The new observations also resulted in the detection of two clumps not seen in the original observations (designated H* and I* in Table \ref{tbl:clump_sizes} and Figure \ref{fig:Ha_maps}). In addition to a change in plate scale for the observations, a new detector on OSIRIS could introduce differences in what clumps were measured.

\begin{table} 
\scriptsize
\begin{center}
\caption{Clump Sizes}
\begin{tabular}{lcccc}
\tableline
\tableline
Clump & Radius & Radius & Luminosity & \sfrd \\
& (mas) & (kpc) & ($\rm 10^{40} \ erg \ s^{-1}$) & ($\rm 10^{-2} \ M_{\sun} \ yr^{-1} \ kpc^{-2}$)\\
\tableline
\multicolumn{5}{c}{2014, 0.1" observations} \\
\tableline
A & 407 & 3.14 & 39.1 & 5.7 \\
B & 143 & 1.10 & 3.54 & 4.1 \\
C & 149 & 1.15 & 2.92 & 3.2 \\
D & 129 & 0.99 & 2.67 & 3.8 \\
E & 128 & 0.99 & 2.04 & 3.0 \\
F & 85 & 0.65  & 1.37 & 4.5 \\
G & 123 & 0.95  & 1.61 & 2.6 \\
\tableline
\multicolumn{5}{c}{2017, 0.05" observations} \\
\tableline
A1 & 272 & 2.10 & 30.2 & 9.8 \\
A2 & 193.5 & 1.49 & 19.3 & 12.4 \\
B & 181 & 1.39 & 13.3 & 9.8 \\
H* & 134 & 1.03 & 4.07 & 5.4 \\
I* & 81.5 & 0.63& 1.61 & 5.9 \\
\tableline
\multicolumn{5}{c}{2017, 0.05" observations binned to 0.1"} \\
\tableline
A & 375 & 2.89 & 52.8 & 9.0 \\
B & 184 & 1.42& 12.3 & 8.8 \\
\tableline
\tableline
\label{tbl:clump_sizes}
\end{tabular}
\tablecomments{clump properties for observations of object 42042481 compared in Figure \ref{fig:Ha_maps}}
\end{center}
\end{table}

In order to determine the reason for the detection of these additional clumps we compare the flux and \sfrd \, of all clumps detected in the new 0.05" observations to those found in the old 0.1" observations as well as the results of binning the 0.05" observations to match the resolution of the 0.1" plate scale. These comparisons are shown in Figure \ref{fig:clump_comparison} (Appendix \ref{app:new_obs}). Since both the flux and \sfrd \, of clumps H* and I* are higher in the 0.05" observations than some of the small clumps in the original observations this cannot be the reason for the detection. Another possible cause for varying detections is the quality of the seeing on each night of observations. In order to investigate this we compare the PSF of the tip-tilt star used for the observations of object 42042481 as well as the seeing measurements from the MASS/DIMM instruments on Mauna Kea. The seeing measurements are reported in Table \ref{tbl:seeing} and the tip-tilt star comparison is shown in Figure \ref{fig:PSF} with widths in kpc denoted by dashed lines in Figure \ref{fig:clump_comparison}. The PSF and seeing across these two nights is very similar and indicates that this also is not the primary cause of detecting new clumps.

It is probable then that these detection differences stem from how we define and find clumps in our analysis. A clump is defined to be a local peak in \ha \, flux which is separated from the next local peak by more than 2 pixels in the \ha \, map \citep{mied16}. All clumps in the 0.05" observations are separated by a distance of more than 4 pixels (2 pixels at the 0.1" scale) from their nearest detected neighbor but are still not detected in the version of the cube binned to match the resolution of the 0.1" observations. These \ha \, peaks are then likely being spread out over more pixels leading to less defined peaks and/or smaller separations between them. The introduction of a new detector between these observations could also reduce the noise in the data leading to an increased SNR (even with the lower \sfrd \, of new clumps H* and I*) and definition between \ha \, peaks, however the difference between the new 0.05" observations before and after being binned to 0.1" plate scale resolution indicates that plate scale is the main driver of the detection differences.

The difference in resolution for these observations results in a similar change in the size of clump A ($\sim1.7\times$ smaller) to that seen by \citet{cav17}, but less than that seen by \citet{fish16} with their degraded images. This difference is likely due to the differences in resolution: \citet{fish16} have a factor of 8 difference in resolution between their local and degraded images, while we have only a factor of 2 difference. It should be noted that our results are only for one galaxy in the sample and while this is an interesting test case it may not be representative of the galaxy population as a whole.

To investigate the possible effect of resolution on the scaling relations determined for a large sample of data we apply the corrections determined by \citet{fish16} to the unlensed, high-redshift data sets. We use these corrections since they are determined for a larger sample of local galaxies. The ``true" correction in fact varies for each study and even each clump based on the resolution achieved. However, exactly what the true correction should be is not yet clear; the three cases discussed here all have different ratios for the change in resolution to the change in clump size. As this ratio is highest for the study by \citet{fish16} we use this correction as the most dramatic change we may expect to see for these samples. This translates to increasing the calculated \sfrd \, by a factor of 6, reducing the measured \lha \, by a factor of 5, and reducing the measured clump radius by a factor of $\sqrt{30}$. This results in a reduced scatter of \sfrd \, at fixed radius (from $\sim$3dex to $\sim$2dex) with the exception of the $z\approx0$ \hii region group \citep{kenn03, gal83, ars88, bas06, roz06, mon07}. 

The same \sfrd \, break as section \ref{sec:SFRDbreak} was applied to this corrected data and the two subsets were fit individually. This resulted in the size-luminosity relation:

\begin{eqnarray}
\begin{array}{cc}
\Sigma_{SFR} > 1 \ \rm M_{\sun} \ yr^{-1} \ kpc^{-2}: L_{H\alpha} \propto r_{clump}^{1.7} \ (corrected) \nonumber\\
\Sigma_{SFR} \leq 1 \ \rm M_{\sun} \ yr^{-1} \ kpc^{-2}: L_{H\alpha} \propto r_{clump}^{2.9} \ (corrected)
\end{array}
\end{eqnarray}

\begin{figure*} [h]
\gridline{\fig{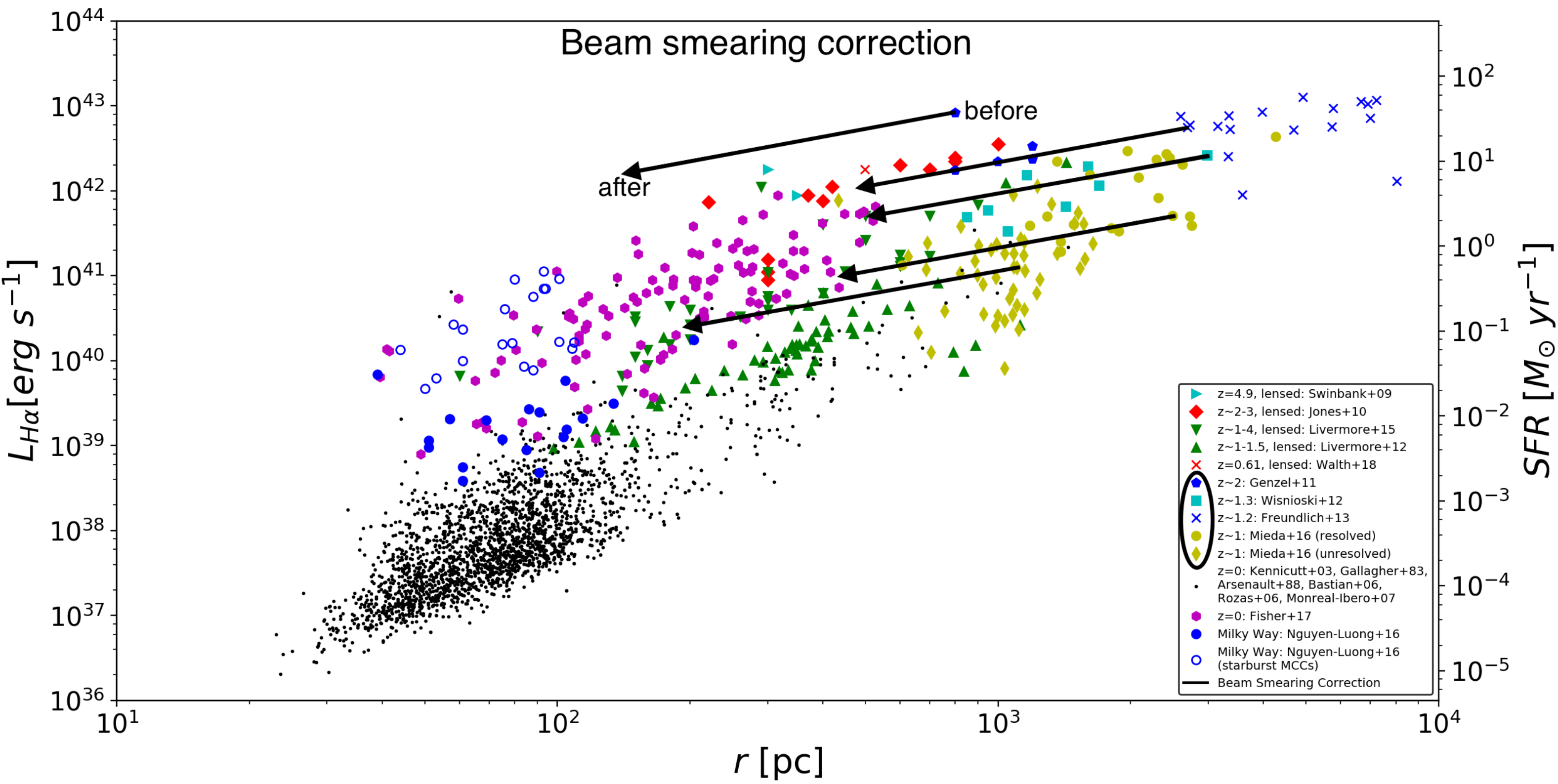}{0.65\textwidth}{(a): Influence of beam smearing}}
\gridline{\fig{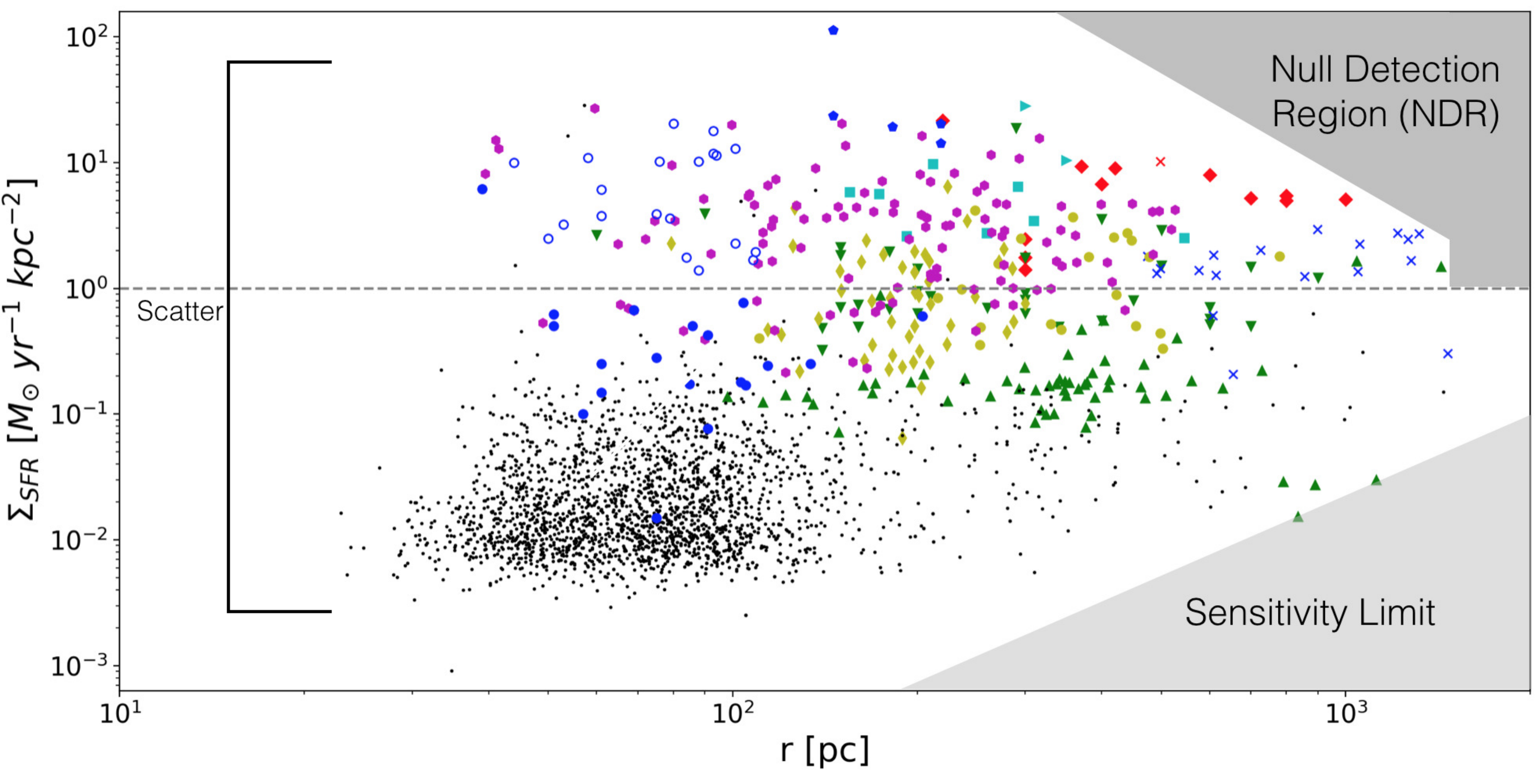}{0.65\textwidth}{(b):\sfrd \, and size after corrections}}
\gridline{\fig{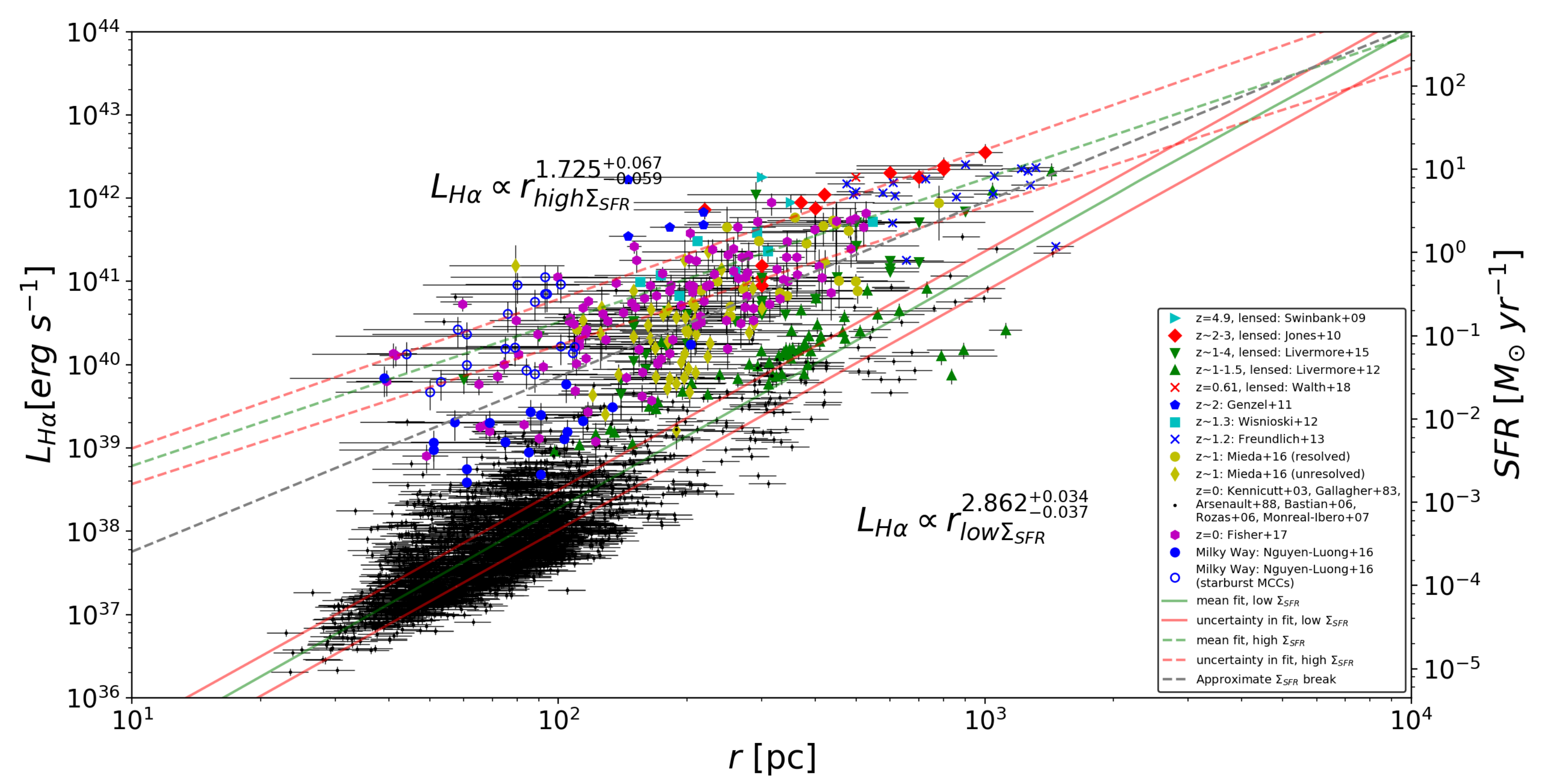}{0.65\textwidth}{(c): Clump size-luminosity relation after correction}}
\caption{(a): Illustration of the beam smearing correction applied to the high-redshift unlensed samples. The correction \protect\citep[from][]{fish16} results in reducing the measured \lha \, by a factor of 5 and the measured clump radius by a factor of $\sqrt{30}$. This in turn increases \sfrd \, by a factor of 6 (moving points down and to the left). Studies which required corrections ($z\gtrsim1$, unlensed) are circled in the legend. (b): Clump size plotted against \sfrd \, with corrections for beam smearing applied. The grey dashed line shows the break at \sfrd \, $\rm = 1 \ M_{\sun} \ yr^{-1} \ kpc^{-2}$ used in dividing the data into high and low \sfrd \, subsets. The shaded region to the lower right shows the regime of data now missing due to instrumental sensitivity limits, while the shaded region in the upper right shows the new ``Null Detection Region" which is not due to a sensitivity limit. (c): Clump size and luminosity relation for the high and low \sfrd \, bins after corrections for beam smearing. The dashed red and green lines show the best fit to the high \sfrd \, data, while the solid red and green lines show the best fit of the low \sfrd \, data. The scaling relations for these two subsets are consistent with those determined before beam smearing corrections indicating that while this moves individual clumps into a different subset, it does not have an impact on the scaling relations within these groups of data. \label{fig:corrected}}
\end{figure*}

Figure \ref{fig:corrected} shows the effect of the beam smearing corrections on the high-redshift unlensed data (a), the application of the \sfrd \, break to the corrected data (b), and the fit to the two sets of data resulting from this break (c). Figure \ref{fig:corrected}b also illustrates the reduction in the influence of the ``Null Detection Region" and sensitivity limit.

Even after applying these corrections to individual clumps, the overall scaling relations of these high and low \sfrd \,  bins does not change significantly. Individual clumps do change bins, but this does not change the overall slope. However, the break at \sfrd \, $\rm = 1 \ M_{\sun} \ yr^{-1} \ kpc^{-2}$ is more clearly evident for large clumps after this correction is applied (Figure \ref{fig:corrected}b compared to Figure \ref{fig:SFRD}).

One caution with this correction is that of the large clumps observed in surveys with lower spatial resolution, it may be that only some of them are actually made up of multiple smaller clumps.  There are clumps of similar size observed in lensed surveys \citep{jon10, liv12, wal18} that have much lower spatial resolution limits, so these large clumps do exist. How much of the population consist of large clumps versus groups of smaller clumps is not yet known, and the effect could be less significant than what is determined here. Due to the uncertain nature of this correction we use uncorrected values for the remainder of this paper.

\subsection{Redshift Evolution} \label{sec:redshift}
Data from all of the studies were grouped by redshift into four bins to investigate whether there is a redshift evolution for the relationship between clump size and luminosity. \citet{liv12, liv15} suggest that the intercept of this relationship does evolve with redshift, but \citet{wisn12} and \citet{mied16} find that their high-redshift samples follow similar scaling relations when including local \hii regions. The bins used here are (i) $z\approx0$ \hii regions (data set \#10 only, as designated in Table \ref{tbl:data_sets}), (ii) all $z\sim0$ (data set \#10-12 in Table \ref{tbl:data_sets}), (iii) $0.6\leq z < 1.5$, and (iv) $z\geq1.5$. The inclusion of two different $z\sim0$ bins is due to the differing nature of the star forming regions of these samples. The data in bin (i) is from various studies of local star forming \hii regions \citep{kenn03, gal83, ars88, bas06, roz06, mon07}, while the second bin includes this data as well as the clumps in \citet{fish16} from low redshift galaxies with turbulent disks and the Milky Way MCCs from \citet{ngu16}. \sfrd \, is higher in these additional clumps and therefore they provide a local analog to the high-redshift galaxies like those in the lensed samples with higher \sfrd; hence the use of two separate low-redshift bins.

\begin{figure*}[]
\epsscale{1.2}
\plotone{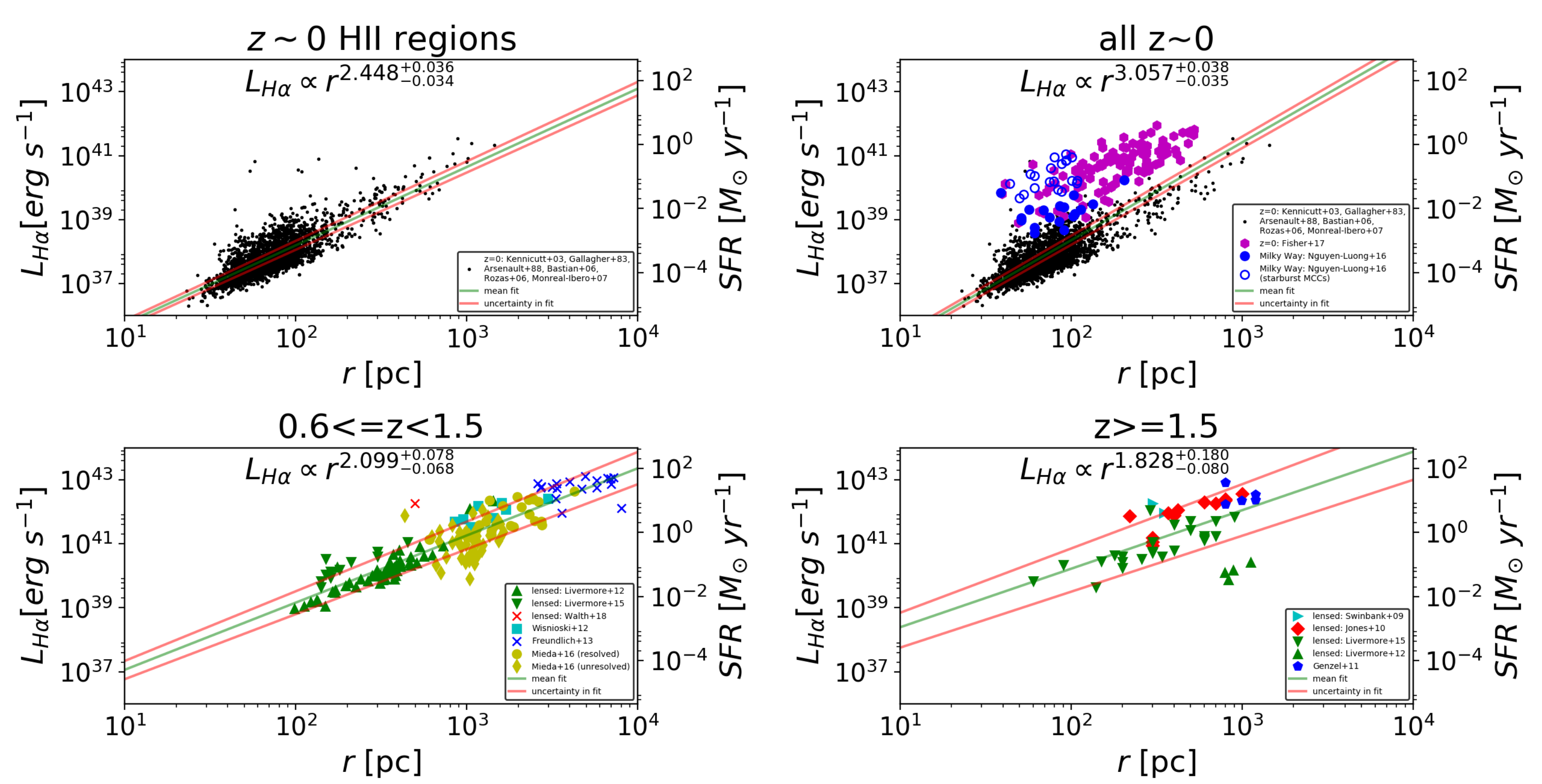} 
\caption{PyStan fit to data from each redshift bin. Upper left: bin (i). Upper right: bin (ii); Lower right: bin (iii); Lower left: bin (iv). The uncertainty on the fit determined is larger for the smaller bins (an order of magnitude for bin(iv)) as there is less data to constrain the fit. The variation seen in the slopes between bins is in part due to variation in the intercept which is not as well constrained with the smaller data sets after binning. However, the posterior probability distribution still appears normal for at least bins (i)-(iii). This leads to a caution on how the data sets are binned and fitted to avoid a case of a well constrained fit that is not physically reasonable or reliable. \label{fig:redshift}}
\end{figure*}

Each bin was fit separately using PyStan configured as discussed in Section \ref{sec:PyStan}, and are presented in Figure \ref{fig:redshift}. In bin (i) \lha $ \sim r ^{2.45^{+0.04}_{-0.03}}$, in bin (ii) \lha $ \sim r ^{3.06^{+0.04}_{-0.04}}$, in bin (iii) \lha $ \sim r ^{2.10^{+0.08}_{-0.07}}$, and in bin (iv) \lha $ \sim r ^{1.83^{+0.18}_{-0.08}}$. This shows that the slope does vary somewhat in each redshift bin, however, this is partly due to the smaller size of the data sets once binned; particularly for the highest redshift bin which only consists of 47 clumps. As can be seen here and in Table \ref{tbl:fit_params2D}, the uncertainty on the slope of bin (iv) is an order of magnitude greater than the other bins which have more data points.  This also leads to a less constrained intercept for bin (iv) which would affect the slope value determined.  Therefore, it is difficult to say for sure whether there is a redshift evolution to the clump size-luminosity scaling relation.

\subsection{Star Formation Dependencies: Gas Fraction and Velocity Dispersion}\label{sec:3D}

As has been shown in the previous sections, the star forming relations of clumps likely do not simply scale with size. There are other properties of the clumps that could influence this relationship and partly account for the large scatter in the data. So far we have used a third parameter, $\Sigma_{SFR}$, to determine a break in the power-law, but the dependence on a third parameter may not be a Heaviside step function, it may be a continuous dependence which needs to be incorporated as an additional dimension to the fit. 

The velocity dispersion of the gas in the clumps gives an indication of the turbulence which likely influences the star formation rate. Here we use this to fit the relationship in Equation \ref{eqn:3D} with $\delta$ being replaced by $\sigma$. The fit converges to a consistent solution of \lha $ \propto r^{1.03} \times \sigma^{2.21}$ with a reasonable posterior probability distribution, indicating that there may be a continuous dependence of the star forming relationships on the velocity dispersion of the clumps. Only some of the clumps used in previous sections have measurements of $\sigma$, reducing the sample size of this fit to 346 of the total 2848 clumps. Fitting these 346 clumps with $\sigma$ included as a third parameter reduces the overall scatter by $\sim92\%$ compared to fitting clump size and luminosity only. The full fit parameters and their uncertainties are shown in Table \ref{tbl:fit_params3DI}.

It has also been suggested that the variations in the size-luminosity relationship determined for different data sets is due to differences in the gas fraction (\fgas) of the star forming regions which may evolve with redshift \citep{liv12, liv15}. In order to test this we again fit a relationship of the same form as Equation \ref{eqn:3D}, replacing $\delta$ with \fgas \, of the host galaxy. Ideally \fgas \, of the individual clumps would be used, but this is currently only known for the host galaxies as a whole and only for 157 of the total 2848 clumps. Further, two of the four samples used here \citep{mied16, wal18} rely on indirect estimates of \fgas \, rather than CO measurements. This data is also from a relatively small subset of the overall sample, but it agrees well with a theoretical dependence of the clump luminosity on both the clump size and the gas fraction. The fit to this data results in a scaling relationship of \lha $\propto r^{1.35} \times f_{gas}^{0.47}$. Interestingly, the scaling for \fgas \, is close to the relationship predicted by Toomre instability (\lha $\propto r^{2} \times$ \fgas$^{0.5}$). Adding \fgas \, as a third parameter also reduces the overall scatter by $\sim78\%$ compared to the 2D size-luminosity fit of these 157 clumps. The full parameters determined in the fit are reported in Table \ref{tbl:fit_params3DII}.  

Like the two-dimensional fit to the clump scaling relations, these multi-parameter fits also give a good fit to the data while spanning the parameter space well. However, these relationships suffer from smaller data sets and we caution against over interpretation of these early results (particularly when it comes to the reduction in scatter). More measurements of \fgas \, and $\sigma$ would aid in further constraining these fits and investigating the relationship for subsets of the overall sample.

\section{Discussion}\label{sec:discussion}

What power-law relationship is determined for the clump size and \lha \, has important implications for the physical processes occurring in the clumps and driving their formation. It is thought that clumps form at regions of gravitational instability in the disk, corresponding with a Toomre parameter $Q < 1$ \citep{toom64, genz11, wisn12}. If the clump or \hii region is represented by a Str\"{o}mgren sphere then there is a well-defined boundary between the ionized and neutral gas. This type of region would have an expected scaling relation of $L_{H\alpha} \propto r^3$. However, if the geometry of this region is non-spherical then a luminosity scaling relation of \lha $\propto r^2$ would be expected. This scaling also results for clumps which are described by the Toomre mass and scale \citep{genz11}. In the following sections we explore both the Str\"{o}mgren sphere and Toomre instability scenarios, in particular how each of these approximations may delineate between the separation of high and low \sfrd \, data sets. 

\subsection{Toomre Instability} \label{sec:toomre}

A common physical explanation for the scaling relationships seen in the high \sfrd \, data comes from investigating the Toomre mass and scale which are representative of a region that forms under the fastest growing mode of Jeans instability \citep{elm09, mur10, genz11}. The Toomre mass and scale ($M_{T}$, $R_{T}$) given in \citet{genz11} are 

\begin{eqnarray}
M_T \propto Q^{-2}a^{-4} (\frac{\sigma_0}{v_c})^2 M_d \label{eqn:Mt}\\
R_T \propto Q^{-1}a^{-2} \frac{\sigma_0}{v_c}R_d \label{eqn:Rt}
\end{eqnarray}
where $R_d$ and $M_d$ are the radius and mass of the disk respectively, $\sigma_0$ is the local velocity dispersion of the gas, $v_c$ is the circular velocity, and $a$ is a constant describing the disk rotation curve. By solving for Q in Equation \ref{eqn:Mt} and substituting into Equation \ref{eqn:Rt}, we arrive at the relationship

\begin{equation}
M_T \sim R_T^2 R_d^{-2}M_d \ or \ M_T \propto R_T^2
\end{equation}

In order to put this in terms of the clump luminosity we turn to an empirical linear relationship locally between the dense gas mass of molecular clouds and their star formation rates \citep{gao04, wu05, lad10}; which also has a theoretical basis in the radiation pressure on \hii regions from star formation. From Equation 13 in \citet{mur10} describing the force due to this radiation pressure, $M_\star \propto L$ in the optically thin limit (optically thin to far-infrared emission while optically thick to ultraviolet). If we are observing clumps that are optically thin to \ha \, emission and assume the Toomre mass traces the dense gas in star forming regions and Toomre scale represents their size, then we expect a clump size-luminosity relationship of \lha $\propto r^2$. This approximates the observed \lha $\propto r^{1.7}$ we find fitting the high \sfrd \,clumps. 

We suspect many clumps (or substructures within) may in fact be optically thick to \ha, making the assumption that $M_\star \propto$ \lha \, tenuous. As a check of this assumption we calculate the estimated virial mass of the clumps which have measurements of velocity dispersion:
\begin{equation}
M_{vir} = \frac{\pi^2 \sigma^2 r}{3G}
\end{equation}
using the measured velocity dispersion, $\sigma$, and radius, r, of the clumps. Comparing this to the observed \ha \, luminosity we find a nearly linear relationship (\lha $ \propto M_{vir} ^{1.07}$). Thus we assume that the mass of the clumps and the dense gas mass are approximately proportional, leading to the \lha $\propto r^2$ relationship.

When the \sfrd \, power-law break is applied at \sfrd \, $\rm = 1 \ M_{\sun} \ yr^{-1} \ kpc^{-2}$, the high \sfrd \, data follow a scaling relationship close to this $r^2$ value for the full sample ($L\propto r^{1.7}$) and the local analogs alone ($L\propto r^{1.5}$). This indicates that these high \sfrd \, clumps could be forming under the fastest mode of Jean's instability. This Toomre mass and scale argument can not however explain the $L\propto r^{2.8}$ scaling found for the low \sfrd \, clumps.

\subsection{Str\"{o}mgren Spheres} \label{sec:stromgren}

Another suggested explanation for the observed scaling relations is that star forming regions at high-redshift form under Jeans collapse at locations of disk instability and are well represented by Str\"{o}mgren spheres \citep{wisn12}. One of the relationships expected from this model of clumps is a size-luminosity scaling of $L \propto r^3$ which comes from equating the recombination rate (left hand side; Equation \ref{eqn:L^3}) and ionization rate, Q, (right hand side) of the hydrogen gas in a spherical region:

\begin{equation}
\frac{4 \pi}{3} R_{strom}^3 \alpha_B n_H^2 x^2 = Q = \frac{L_{H\alpha}\lambda_{H\alpha}}{h c}
\end{equation}
where $R_{strom}$ is the Str\"{o}mgren radius, $\alpha_B$ is the Case-B recombination coefficient \citep{ost89}, and $n_H$ is the number density of hydrogen atoms.  $x$ is the ratio of free electrons to hydrogen atoms ($x=\frac{n_e}{n_H}$) and is approximately equal to 1 for a fully ionized region. This results in the final size-luminosity relationship of:
\begin{equation}
L_{H\alpha}=\frac{4 \pi h c \alpha_B n_H^2}{3 \lambda_{H\alpha}} R_{strom}^3 \label{eqn:L^3}
\end{equation}

With the clump radii being representative of the Str\"{o}mgren radius, this results in the $L\sim r^3$ scaling for the \ha \, luminosity of the clumps we find when fitting the data set as a whole, but this fit is likely skewed by the large scatter in the overall data set.  However, when the power-law break is applied at \sfrd \, $\rm = 1 \ M_{\sun} \ yr^{-1} \ kpc^{-2}$, the scaling determined for the low \sfrd \, subset is very close to this theoretical relationship at $L\propto r^{2.8}$ for the full data set and $L\propto r^{2.7}$ for just the local analogs. \citet{wisn12} find a relationship of $L \propto r^{2.72\pm0.04}$ for their full data set, and they suggest the shallower slope may be due to the clumps being density bound rather than being idealized Str\"{o}mgren spheres. This would mean that the hydrogen atoms in the star-forming region can recombine faster than they are being ionized. This idea of having density bound clumps is discussed in more detail in \citet{wisn12} and \citet{beck00}.

This does not however explain the $L\propto r^{1.7}$ and $L\propto r^{1.5}$ scaling we see in the high \sfrd \, data for both the full sample and local analogs. A possible explanation for this is that the clump ``radius" is set by the optical depth unity surface, but if the rate of production of ionizing photons $Q$ is large enough then that the surface may not approximate a sphere, i.e.,

\begin{eqnarray}
& R_{strom}(Q) > H, \label{eqn:R>H} \\
& with \ Q=\frac{L_{H\alpha} \lambda_{H\alpha}}{h c} f_{rec} \label{eqn:Q_factor} \\
and \ & f_{rec} \approx \left({H\over R_{\rm S}(Q)}\right)
{3\over 2}\left[1-{1\over 3}\left({H\over R_{\rm S}(Q) }\right)^2\right]    \label{eqn:ionization_fraction}
\end{eqnarray}
where H is the scale height of the disk and $f_{rec}=1$ for an ideal Str\"{o}mgren sphere (giving the scaling in Equation \ref{eqn:L^3}).

Plugging Equation \ref{eqn:Q_factor} into Equation \ref{eqn:L^3} we get the Str\"{o}mgren radius as a function of Q:

\begin{equation}
R_{strom}(Q) = \left[\frac{3Q}{4 \pi \alpha_B n_H^2}\right]^{1/3} \label{eqn:R_s}
\end{equation}

Combining this with Equations \ref{eqn:ionization_fraction} and \ref{eqn:Q_factor} (to get back to \lha) we find that for a clump with radius greater than the disk scale height,
\begin{equation}
L_{H\alpha} = 2 \pi \alpha_B \frac{h c}{\lambda_{H\alpha}} R_{strom}^2 H \left[1-\frac{1}{3}\left(\frac{H}{R_{strom}(Q)}\right)^2\right]
\end{equation}

When $R_{strom}(Q) > H$ the term in brackets on the right is approximately 1, giving the scaling:
\begin{equation}
L_{H\alpha} \sim R_{strom}^2 H
\end{equation}

Since we only plot the \ha \, luminosity against clump size, this gives us the nearly $L \sim r^2$ scaling seen in the high \sfrd \, data sets and could explain the reason for a power-law break. Note that this would give $L_{H\alpha} \propto r^3$ for cases where $R_{strom} \approx H$.

As a check of the power-law break we use, we can calculate the critical \ha \, luminosity, $L_{H\alpha,crit}$ above which we would expect to see $L_{H\alpha} \propto R_{strom}^2$. This critical point would be where $R_{strom}(Q) \approx H$, with H:
\begin{equation}
H=\frac{\sigma}{v_c}R_g \label{eqn:scale_height}
\end{equation}
where $R_g$ is the galactocentric radius, $\sigma$ is the velocity dispersion of gas in the disk (which for the largest clumps in the Milky Way is similar to the velocity dispersion of the clump), and $v_c$ is the circular velocity of the disk.

Combining this with Equation \ref{eqn:R_s} we arrive at an expression for the critical luminosity at which the scaling would switch from $r^3$ to $r^2$:
\begin{equation}
L_{H\alpha,crit} \approx \frac{4 \pi h c \alpha_B n_H^2}{3 \lambda_{H\alpha}} \left(\frac{\sigma}{v_c}\right)^3 R_g^3 \label{eqn:L_crit}
\end{equation}

If we take an average clump with a velocity dispersion, $\sigma=50$ $\rm km s^{-1}$, density, $n_H=10$ $\rm cm^{-3}$, disk circular velocity, $v_c =250$ $\rm km s^{-1}$, and galactocentric radius, $R_g=1$ kpc, we arrive at a value of \lha $\rm \approx 2\times10^{40} erg \ s^{-1}$ and a scale height $H=0.2$ kpc. Comparing with the size luminosity plot in Figure \ref{fig:SFRD_break}, this is approximately where the \sfrd \, cut-off lies for a clump radius of $0.2$ kpc.

\begin{figure*}[]
\epsscale{1.0}
\plotone{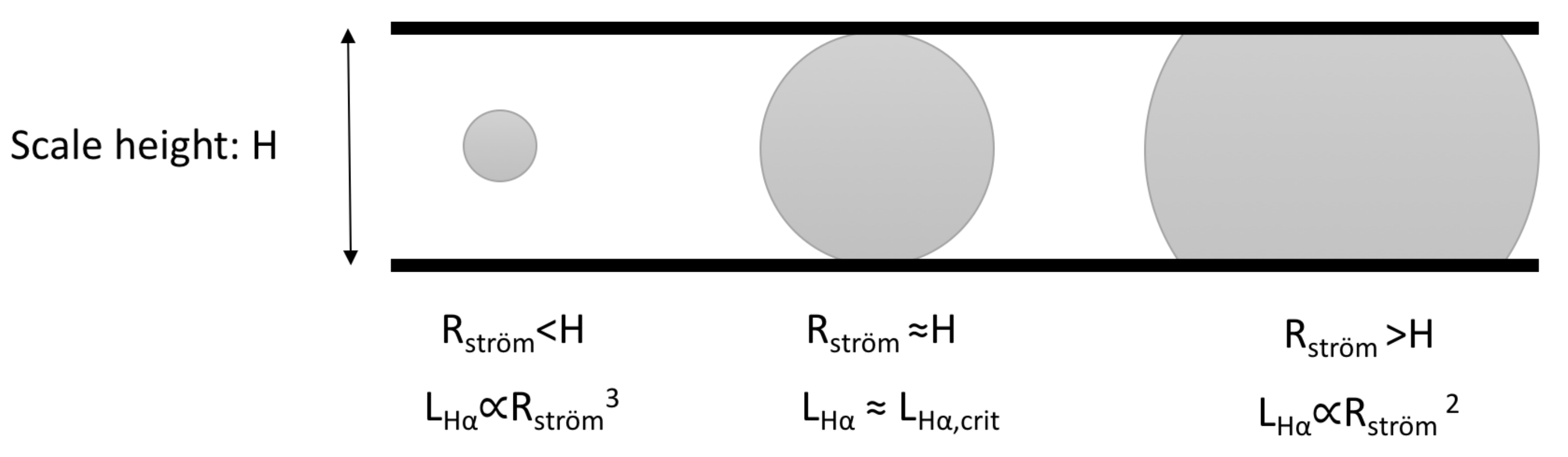} 
\caption{Illustration of the three regimes of clump size relative to the scale height of the host galaxy disk. The set of clumps smaller than their host galaxy scale height ($R_{strom} < H$) would give an expected size-luminosity scaling of \lha $\propto r^3$. Those with radii equal to the host galaxy scale height would fall along the critical luminosity and clumps with radii larger than the disk scale height would have an expected \lha $\propto r^2$ scaling.  \label{fig:scale_height}}
\end{figure*}

Figure \ref{fig:scale_height} illustrates the physical difference and differences in the size-luminosity scaling relationship expected for clumps in these three size regimes relative to the scale height of the disk: $R_{strom} < H$, $R_{strom} = H$, and $R_{strom} > H$.

\subsection{Bound vs. Unbound Clumps}\label{sec:sigma_break}
When investigating local star forming GMCs and MCCs, \citet{ngu16} introduce a virial parameter based on the velocity dispersion ($\sigma$) which divides between gravitationally bound and unbound star forming regions. The velocity dispersion is a measure of the turbulance in the clumps and is used here as an indication of whether or not the clumps are gravitationally bound; a property that may cause a variation in the star formation scaling relations.  

To investigate this we would like to introduce a ``break" in the power-law that is dependent on $\sigma$ of the ionized gas in the clumps. However, the velocity dispersion was only measured for a smaller number of data sets resulting in a large scatter. Fitting these small samples results in a poorly constrained fit and uncertainties which are on the same order as the nominal value (an order of magnitude greater than the uncertainties on larger samples).  Without a larger data set to base these fits on it is difficult to say whether there is a break in the scaling relations based on the velocity dispersion cut-off. The influence of the clump velocity dispersion was still able to be investigated in Section \ref{sec:3D} as a third fitting parameter since the sample size did not suffer from being divided into two subsets. We believe it is important for studies to include the velocity dispersion of the individual clumps in future investigations.
\newpage
\subsection{Feedback} \label{sec:feedback}
In addition to providing evidence for two different clump populations, Figure \ref{fig:SFRD} also provides valuable information about these populations from what we \textit{do not} observe. There is a lack of star forming regions with both large size and high \sfrd: the region with \sfrd $ > 1 M_{\sun} \ yr^{-1} \ kpc^{-2}$ and $r > 10^3$ pc. Due to this corresponding to clumps which would be both large and have high surface brightness, the lack of observations in this region cannot be due to a sensitivity limit. Instead, it is probable that these clumps just do not exist on a timescale which would make them likely to be observed, indicating some type of feedback mechanism regulating these star forming regions.

There are numerous possible feedback mechanisms put forward for disrupting star forming regions including supernovae explosions, jets due to star formation, thermal pressure from ionized gas, and radiation pressure due to dust absorbing and scattering photons \citep[e.g.][]{mur10,fall10}. \citet{mur10} investigate these factors in detail and how they influence a wide range of star forming regions including GMCs in the Milky Way and clumps seen in a $z \sim 2$ galaxy. They find that in all cases the earliest supernovae would occur after the star forming region was already being disrupted and therefore could not be the main factor.  The jets are also shown to only be a main factor early in the disruption of the star forming region while the thermal pressure is important in the Milky Way GMCs but not in the more luminous star forming regions like the $z \sim 2$ clumps. The radiation pressure is found to be the dominant feedback mechanism contributing to the disruption of star forming regions \citep[also found by ][]{fall10}. However, more recent simulations by \citet{krum12, krum13} show that radiation trapping is negligible in giant clumps since it destabalizes the outflow winds. \citet{dek13} argue that this means that steady winds from radiation pressure would not disrupt the clumps before they migrate to the disk center. 

\citet{man17} discuss the two main scenarios seen in simulations for the lifetimes of clumps at high-redshift. For simulations which only include supernova feedback the clumps are not disrupted and migrate to the center of the disk to form and grow the bulge on an orbital timescale (250-500 Myr). However, in simulations that include radiation pressure feedback clumps tend to be disrupted on a dynamical timescale (50-100 Myr). In an investigation of a massive galaxy between $z\sim2.2 - 1$ using the FIRE simulations (including radiation pressure and other forms of stellar feedback) the average lifetime of clumps above $\rm 10^8 \, M_{\sun}$ is found to be comparatively short at $\sim$22 Myr \citep{okl17}. For the clumps included in this study which have measurements of velocity dispersion we find an average dynamical time of 3.5 Myr. While the mechanisms of feedback in high-redshift clumps may not be fully understood it is possible that disruption of local and high-redshift clumps are leading to the lack of large, high \sfrd \, clumps observed.

\subsection{Possible Sources of Bias} \label{sec:bias}
By combining different data sets (i.e. lensed and unlensed, high-redshift and low-redshift) there are various selection biases from each survey, which may have an impact on our results, especially when we split the data into smaller subsets. The unlensed surveys typically probe more massive galaxies than the lensed surveys which could introduce differences in the clumps present, however, we do not find significant differences in the scaling relationships of data from these two types of surveys. These selection effects still do weight the overall data set at high-redshift more heavily towards massive galaxies which are easier to observe (Figure \ref{fig:histograms}). The unlensed surveys at $z>1$ tend to have higher limits to the spatial resolution which introduces the possibility of some of the observed clumps actually being complexes of smaller clumps whose properties are more similar to those observed in the lensed surveys, which still results in much the same scaling relations albeit with different scatter. However, it is unknown at this point what percentage of the clump population observed in these unlensed high-redshift surveys may actually be clump complexes since clumps of similar radius are also observed in lensed surveys with lower resolution limits \citep{jon10, liv12, wal18}.  While gravitational lensing provides the opportunity for better spatial resolution it should be noted that there are larger uncertainties involved (particularly in spatial measurements) due to the lensing model. In order to test the influence of these selection effects we fit the lensed and unlensed high-redshift samples individually (with no local analogs; Appendix \ref{sec:lensing}). The slopes between these two fits are consistent within the uncertainties at around $L \propto r^2$ indicating that these two sample types follow the same scaling relationship.

The results of fitting the overall data set is largely influenced by the group of $z\approx0$ \hii regions since this provides many more data points than the high-redshift samples. This is fine if the physical process and scaling relations are the same for these samples, but  as discussed in Section \ref{sec:SFRDbreak} there is evidence that these \hii regions are not the best local analog due to the lower \sfrd \, than the high-redshift star forming regions (part of this difference is of course due to a sensitivity limit at high-redshift) and the inclusion in the full data set creates a large scatter. This scatter results in very different scaling relations when fitting with and without these $z\approx0$ objects ($\sim r^3$ and $\sim r^2$ respectively), so resolving this issue would be highly beneficial in determining the processes occurring in clump formation at high-redshift. Better spatial resolution and surface brightness sensitivity of the more massive galaxies typical of the unlensed sample may help resolve this since these are currently the galaxies that tend to have similar measured \sfrd \, as the local \hii regions. However, if the beam smearing effects discussed in Section \ref{sec:blurring} and \citet{fish16} are important then these galaxies will typically also have much higher intrinsic \sfrd \, than what we are currently measuring causing them to be offset from the $z\approx0$ \hii regions. If, on the other hand, there is a sensitivity limit causing us to currently miss clumps with lower SFR and \lha \, there may be lower luminosity clumps at the same size scales as our high-redshift unlensed samples.

The absence of large clumps characteristic of the high-reshift samples which also have very low \sfrd \, (lower right region of Figure \ref{fig:SFRD}) similar to the $z\sim0$ \hii regions indicates that such a sensitivity limit is likely affecting our observations and fitting. In particular this may be forcing the slope of the low \sfrd \, subset to a higher value, closer to $L\propto r^3$. To investigate this we calculated the observed flux density which corresponds to clumps with these lower values of \sfrd \, at $z\sim1$ and $z\sim2$ (illustrated in Figure \ref{fig:sensitivity} in Appendix \ref{app:figures}). The actual sensitivity limit for each instrument will be dependent on the configuration and will vary with the performance of the AO system (if one was used). The lack of large, low \sfrd \, clumps in Figure \ref{fig:SFRD} indicates that such a limit is impacting the clump population being observed.

Differences in how extinction was accounted for between samples can introduce an additional source of bias in our investigation. Not accounting for the effects of extinction in the determination of \lha \, may cause some clumps to be artificially shifted down on the size-luminosity plots. There are some studies used here that do not take this into account for their measurements. Among the lensed samples, \citet{swin09, jon10} do not account for extinction effects, while \citet{liv12, liv15, wal18} correct for the average extinction in each galaxy. Among the unlensed samples, \citet{genz11, freun13, mied16, fish16} correct for the average extinction while \citet{wisn12} do not apply a correction. Among the \hii regions \citet{kenn03, bas06, mon07} correct for the extinction of individual star forming regions, \citet{ars88} contains some objects corrected for average host galaxy extinction and others uncorrected, and \citet{gal83, roz06} do not apply a correction. In \citet{mied16} (z$\sim$1, unlensed) correcting for extinction resulted in an average increase in \lha \, by a factor of $\sim2$. Figure \ref{fig:extinction} in Appendix \ref{app:figures} illustrates the effect of adding this average correction to studies which had not accounted for it. Due to where the data from these studies fall on the size luminosity plot this results in an increase in scatter at large clump sizes and the same decrease at small sizes (0.8 dex for a conservative $\rm A_v = 2mag$ correction). This is a relatively small effect and likely does not significantly change the scaling relations we determine.

\hii regions measured in the SMC \& LMC \citep{kenn86}, IC10 \citep{hodg89}, and NGC 6822 \citep{hodg89b} were not included in this analysis, but were plotted with the data and scaling relationships determined here to check that the fits are physically realistic at lower size scales. These star forming regions have similar \sfrd \, to the \hii regions used throughout this paper but still appear to follow the $\sim r^2$ scaling of the fit without the $z\approx0$ \hii regions just with the intercept shifted down.

It should be noted that some of the values measured at lower size scales could be affected by stochastic sampling of the stellar initial mass function (IMF) of the clump regions.  In simulations performed by \citet{cal12} a lower limit on size of 200 pc is used to avoid these effects by keeping the SFR above $\rm 1.3 \times 10^{-3} M_{\sun} \ yr^{-1}$.  Below this limit they report that stochastic sampling of the IMF would have an impact on measurements of SFR indicators like \lha. This SFR limit corresponds to \lha $\rm \sim 3\times10^{38} ergs \ s^{-1}$ with a Chabrier IMF \citep{chab03}, a value which some of the data used in this study does fall below--particularly among the $z\approx0$ \hii regions.  This may add to the uncertainty in the measurements of lower luminosity star forming clumps, but is not likely to have a significant effect on the results using the high-redshift data of this study.

\section{Summary/Conclusion}\label{sec:conclusion}

We compiled a comprehensive set of data on the sizes and luminosities of both local and distant resolved star forming regions from the literature. These data sets were carefully binned based on differences in surveys and clump properties to exhaustively explore potential size-luminosity scaling relationships using MCMC fitting with PyStan. We find the following trends and conclusions from this analysis:

\begin{enumerate}

\item There is a large scatter of order 4 dex in luminosity for a given clump or \hii region size. This scatter may significantly impact the inferred size-luminosity scaling relationship, depending on the choice of sample used in the fit. For example, if the local star forming data from \citet{fish16} and \citet{ngu16} are used then the scaling relation determined is $L\propto r^2$. If the set of $z\approx0$ \hii regions are also included then the scaling relationship becomes $L\propto r^3$. 

\item We observe a break in the size-luminosity scaling relation based on the measured clump \sfrd \, at $\rm 1 \ M_{\sun} \ yr^{-1} \ kpc^{-2}$. Clumps with lower \sfrd \, tend to have luminosities that scale closer to $\sim r^3$, while clumps with higher \sfrd \, tend to have luminosities that scale with $\sim r^2$. This is true for both the low-redshift sample and the entire collated data set.

\item We find that the $L \propto r^3$ scaling can be explained by clumps that are well represented by Str\"{o}mgren spheres which are smaller than the scale height of the disk. We find that if the Str\"{o}mgren radius is larger than the scale height of the disk and some ionizing photons are escaping, then the non-spherical geometry may result in a $L \propto r^2$ scaling. Alternatively, star formation regions driven by Toomre instability may result in a $\sim r^2$ scaling of the high \sfrd \, clumps, but is unable to be extended the low \sfrd \, clumps to yield a $\sim r^3$ scaling.

\item If there exists a power-law break in the size-luminosity scaling relationship of star forming regions, this may indicate a secondary dependence on additional clump properties. We investigated the dependence of the size-luminosity relationship with respect to the host galaxy gas fraction (\fgas) and clump velocity dispersion ($\sigma$), but further data on these parameters are still needed to do a thorough investigation. Additional IFS studies would provide kinematics for galaxies and clumps, while ALMA observations of molecular gas would provide accurate gas fractions for host galaxies and individual clumps. 

\item Spatial resolution effects observed for high-redshift (unlensed) galaxies may alter the measured properties ($r_{H\alpha}$, $L_{H\alpha}$, $\Sigma_{SFR}$) of the clumps. If such beam smearing effects are wide-spread then this could result in an increased artificial scatter, but does not influence the scaling relation results from the applied power-law break at \sfrd $\rm = 1 M_{\sun} \ yr^{-1} \ kpc^{-2}$.

\item We find no evidence for redshift evolution of the clump size-luminosity relation, but more data at higher-redshift bins are still needed. The differences in slopes between redshift bins can not be separated from the potential effects of the small sample sizes and larger uncertainties at high-redshift.

\item We find a scaling relation $L \propto r^2$ for both high-redshift lensed and unlensed clump data sets that are consistent within the uncertainties. Yet we point out that these are still small data sets that should be expanded for further investigation, in particular the high-resolution lensed sample. \newline \newline

\end{enumerate}

\acknowledgments
The authors wish to thank Randy Campbell and Jim Lyke for their assistance at the telescope to acquire the Keck OSIRIS data sets. We appreciate the valuable discussions with Dusan Keres and Karin Sandstrom. The data presented herein were obtained at the W.M. Keck Observatory, which is operated as a scientific partnership among the California Institute of Technology, the University of California and the National Aeronautics and Space Administration. The Observatory was made possible by the generous financial support of the W.M. Keck Foundation. The authors wish to recognize and acknowledge the very significant cultural role and reverence that the summit of Maunakea has always had within the indigenous Hawaiian community. We are most fortunate to have the opportunity to conduct observations from this precious mountain. This research has made use of the NASA/IPAC Extragalactic Database (NED) which is operated by the Jet Propulsion Laboratory, California Institute of Technology, under contract with the National Aeronautics and Space Administration.

\vspace{5mm}
\facilities{Keck:I (OSIRIS)}

\software{PyStan \citep{stan17}, OSIRIS Data Reduction Pipeline:\citep{OSIRIS_DRP}, Matplotlib \citep{matplotlib}}

\appendix

\section{Lensed vs. Unlensed Observations}\label{sec:lensing}
We have binned the data sets into gravitationally lensed and unlensed  high-redshift observations. This was done to test for any influences of selection biases in the data that is typically gathered from lensed versus unlensed surveys at higher redshift. Lensed surveys provide enhanced spatial resolution and can allow us to extend our analysis to lower luminosity galaxies due to the magnification effects \citep{liv15}, which results in the tendency towards lower mass galaxies than can be probed by unlensed surveys. However, the lensing model does introduce larger uncertainties on the measured values, particularly when it comes to the size of clumps.

As is shown in Figure \ref{fig:lensed} there is a very slight difference between the nominal slope values of the lensed (\lha$\sim r ^{2.10}$) and unlensed (\lha$\sim r ^{2.27}$) fits, however these values are consistent within the uncertainties.  This indicates that regardless of the selection differences between the two types of studies, the scaling relations determined from each are consistent. The small offset seen in the intercept between these two bins could then be caused by the effect of beam smearing on the measurements of clump size and luminosity.

\setcounter{figure}{0}   
\renewcommand{\thefigure}{\Alph{section}\arabic{figure}}  

\begin{figure*}[h!]  
\epsscale{2.0}
\plottwo{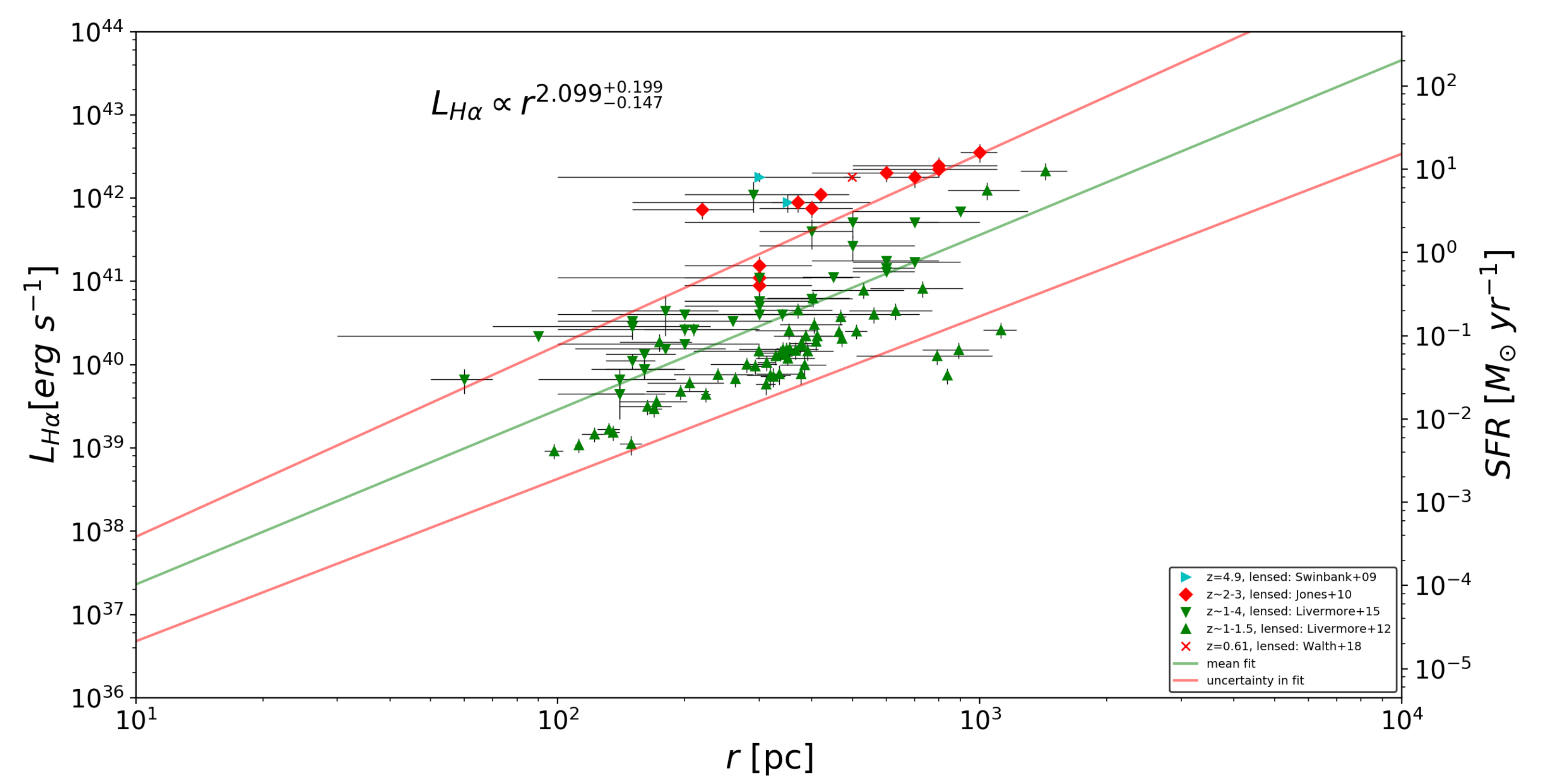}{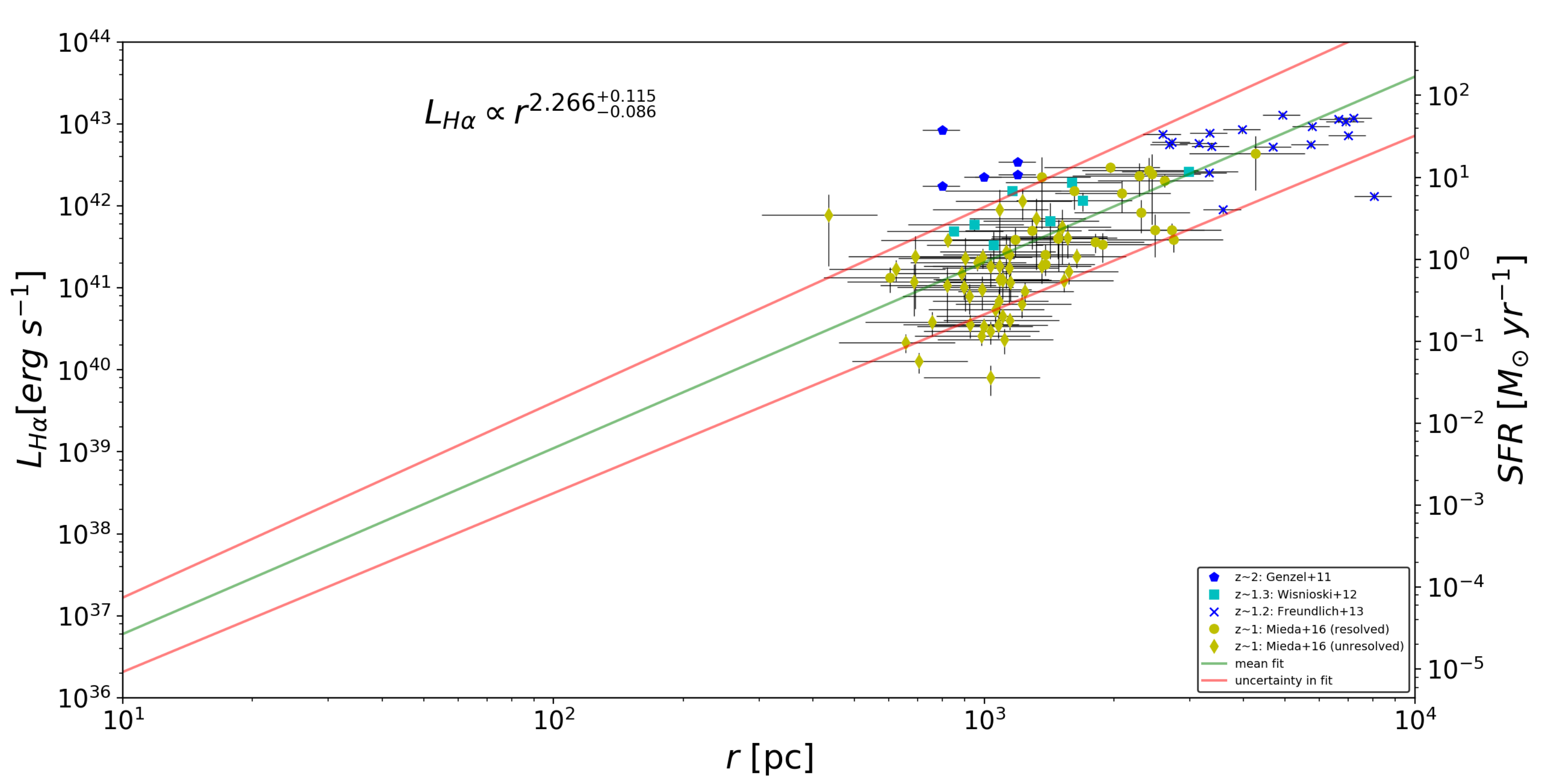}
\caption{Clump size and luminosity relation for high-redshift lensed data sets (top) and unlensed data sets (bottom). Error bars are shown to illustrate the large variations in each data set. NOTE: The $z\approx0$ \hii regions and other local analogs are excluded here since the large number of data points has an overwhelming influence on the fitting. \label{fig:lensed}}
\end{figure*}

\clearpage
\section{Beam Smearing Investigation with Object 42042481}\label{app:new_obs}
\setcounter{figure}{0} 
\setcounter{table}{0}
\renewcommand{\thetable}{\Alph{section}\arabic{table}}
In the investigation of beam smearing effects on measured clump properties we re-observed one of the brightest galaxies of the IROCKS sample \citep[unlensed, $z\sim1$ galaxy 42042481,][]{mied16} at a plate scale of 0.05" in order to increase the resolution over the initial 0.1" plate scale. This resulted in a factor of 2 improvement in spatial resolution (from $\sim$800\,pc to $\sim$400\,pc) and the largest clump breaking into two clumps nearly half the size originally measured (Table \ref{tbl:clump_sizes}). In addition to this clump breaking into smaller components, two new clumps (H* and I*) were also detected in the 0.05" observations. In order to determine the cause of these additional clump detections we compare the flux and \sfrd \, of the clumps detected in the new 0.05" observation (binned to 0.1" resolution and un-binned) with the clumps found in the previous 0.1" observations. This is shown in Figure \ref{fig:clump_comparison} with clumps H* and I* having higher flux and \sfrd \, than some previously detected clumps. This therefore is not the driver of the new detections. The quality of seeing on each night of observations could also lead to differences in clump detection. Therefore both seeing measurements from the MASS/DIMM instruments on Mauna Kea (Table \ref{tbl:seeing}) and the PSF of the tip-tilt star used for each observation (Figure \ref{fig:PSF}) are compared. Both the PSF and seeing measurements across the two nights is similar, suggesting this is not the cause of the new detections either and it is likely in our definition of \ha \, clumps.

\begin{figure*}[h!]
\epsscale{1.7}
\plottwo{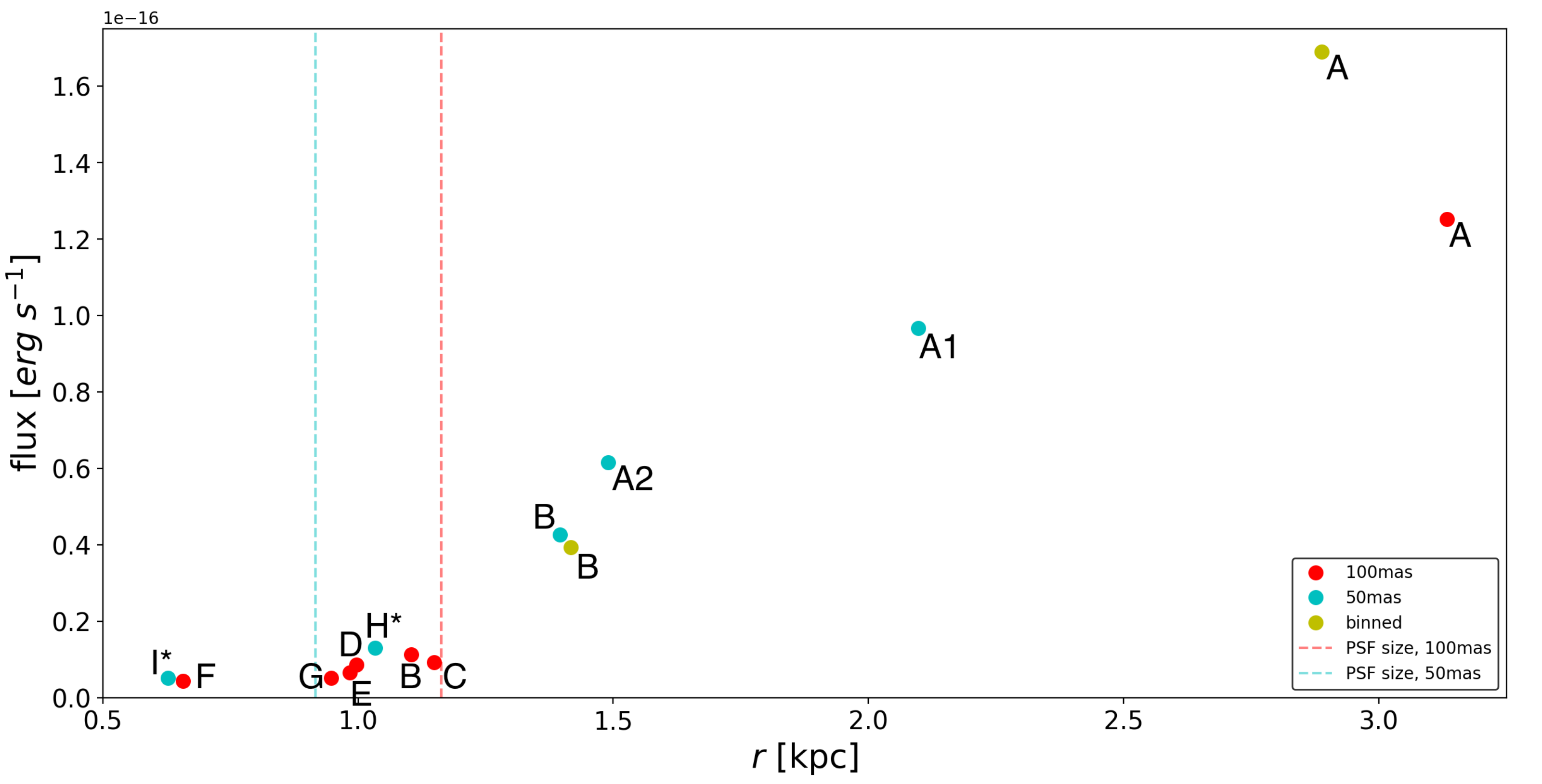}{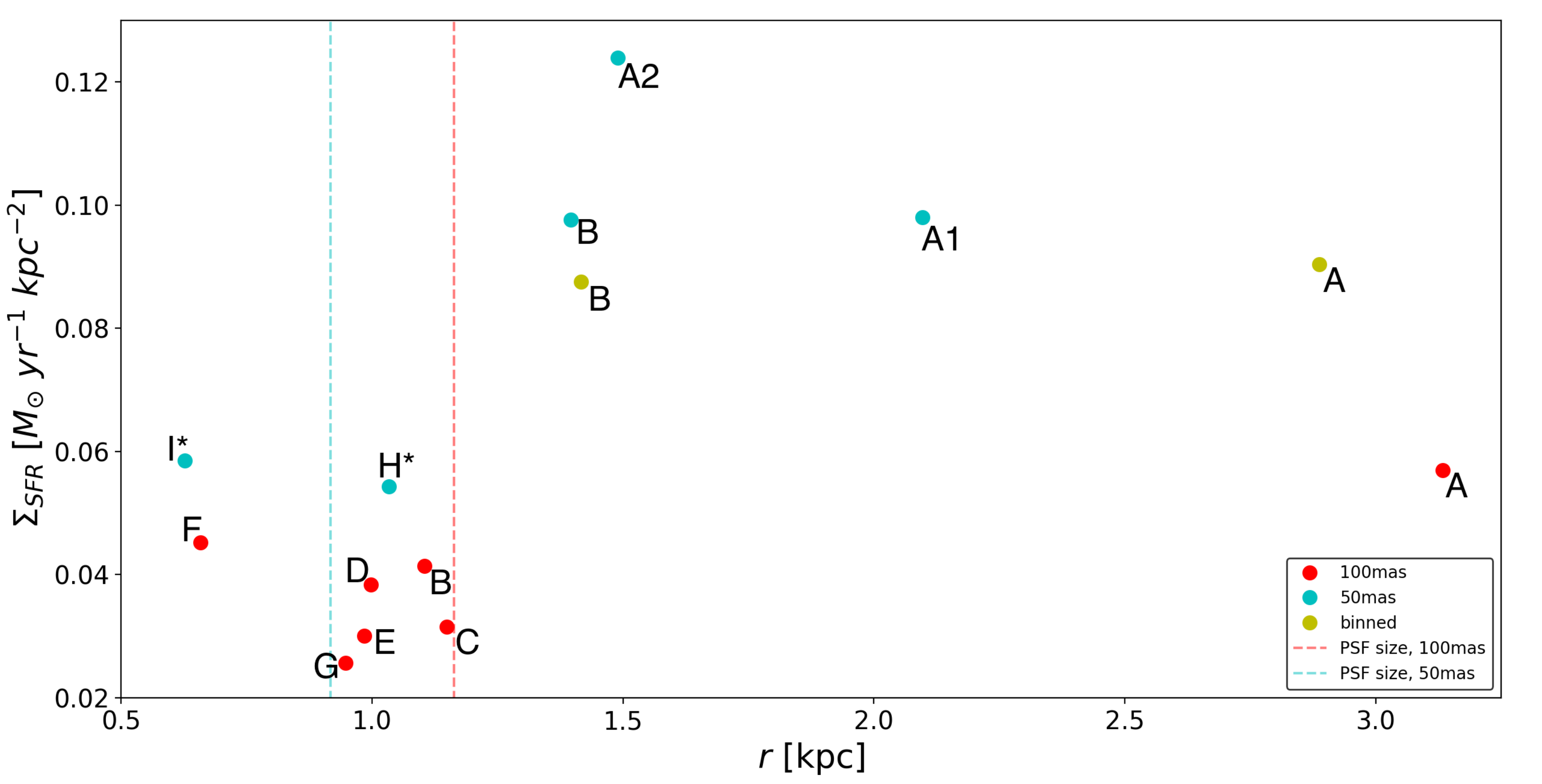}
\caption{Comparison of the flux (top) and \sfrd \, of each clump detected in the 2014 0.1" observations, 2017 0.05" observations,  and the 2017 0.05" observations binned down to 0.1" resolution. The dashed lines show the scale of the PSF for each night and plate scale of observations. The clumps which fall to the left of these lines would be considered unresolved. \label{fig:clump_comparison}}
\end{figure*}

\begin{table}[h] 
\begin{center}
\caption{MASS/DIMM Seeing Measurements}
\begin{tabular}{lcccc}
\tableline
\tableline
Instrument & Mean Seeing & Min Seeing & Max Seeing & Standard Deviation\\ 
& (arcsec) & (arcsec) & (arcsec) & (arcsec)\\
\tableline
\multicolumn{5}{c}{2014 November 8-9; 0.1" observations} \\
\tableline
DIMM & 0.46 & 0.27 & 0.93 & 0.11 \\
MASS & 0.20 & 0.06 & 0.62 & 0.11 \\
\tableline
\multicolumn{5}{c}{2017 August 11-12; 0.05" observations} \\
\tableline
DIMM & 0.58 & 0.30 & 1.59 & 0.18 \\
MASS & 0.23 & 0.06 & 0.52 & 0.09 \\
\tableline
\tableline
\label{tbl:seeing}
\end{tabular}
\end{center}
\end{table}

\begin{figure*}[h!]
\epsscale{0.9}
\plotone{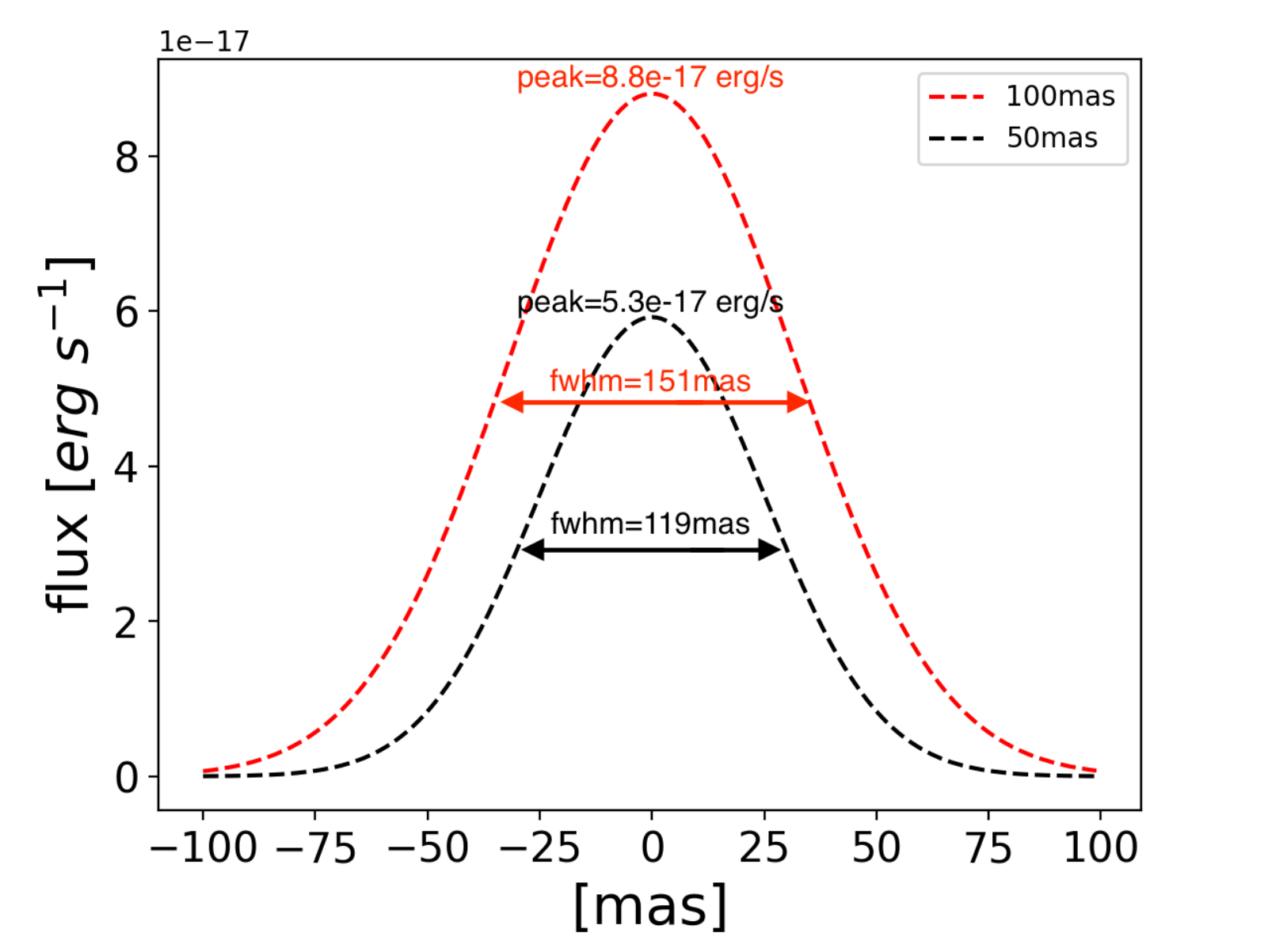}
\caption{Comparison of the tip-tilt star PSF for the 2014 and 2017 observations at a plate scale of 0.1" and 0.05" respectively. The smaller width of the 0.05" PSF could be a consequence of the lower peak flux as the width is smaller by a factor of $\sim0.8$ and the peak flux is lower by a factor of $\sim0.6$. \label{fig:PSF}}
\end{figure*}

\clearpage
\section{Additional Figures} \label{app:figures}
\setcounter{figure}{0}  
The illustration in Figure \ref{fig:sensitivity} shows where sensitivity limits may lie on the size-\sfrd \, plot at different redshifts in this data set. The actual sensitivity limit of each study will vary widely based on the telescope/instrument used and will even vary within studies based on lensing effects. To simplify this we only show the observed flux needed to detect a clump at different levels of \sfrd \, at $z=1$ and $z=2$.

\begin{figure*}[h!]
\epsscale{0.9}
\plotone{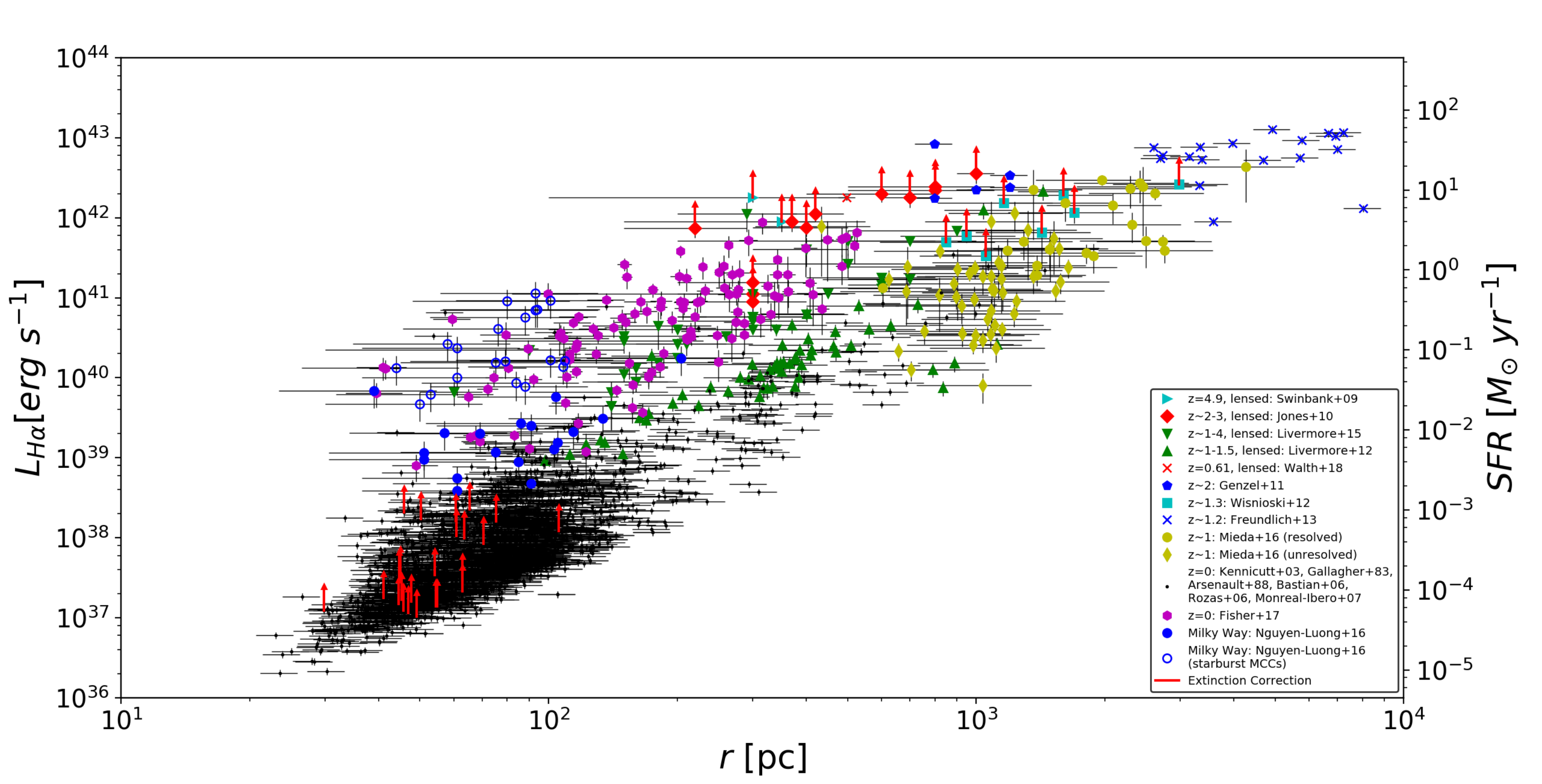}
\caption{Clump size and luminosity for all the data used throughout this paper with illustration of the estimated influence of adding extinction correction to those samples which do not already do this (data set \#'s 1,2,7,10,12 from Table \ref{tbl:data_sets}). The correction used is the average of the affect observed in \protect\citet{mied16} of increasing \lha \, by a factor of $\sim2$. The correction is only shown for a few of the $z\approx0$ \hii regions (data set 10) but would apply to all.  \label{fig:extinction}}
\end{figure*}

Figure \ref{fig:extinction} illustrates the estimated influence of adding extinction corrections to the luminosities measured in studies which did not already include these corrections. For these studies an average 2$\times$ increase in \lha \, would be expected.

\begin{figure*}[h!]
\epsscale{0.9}
\plotone{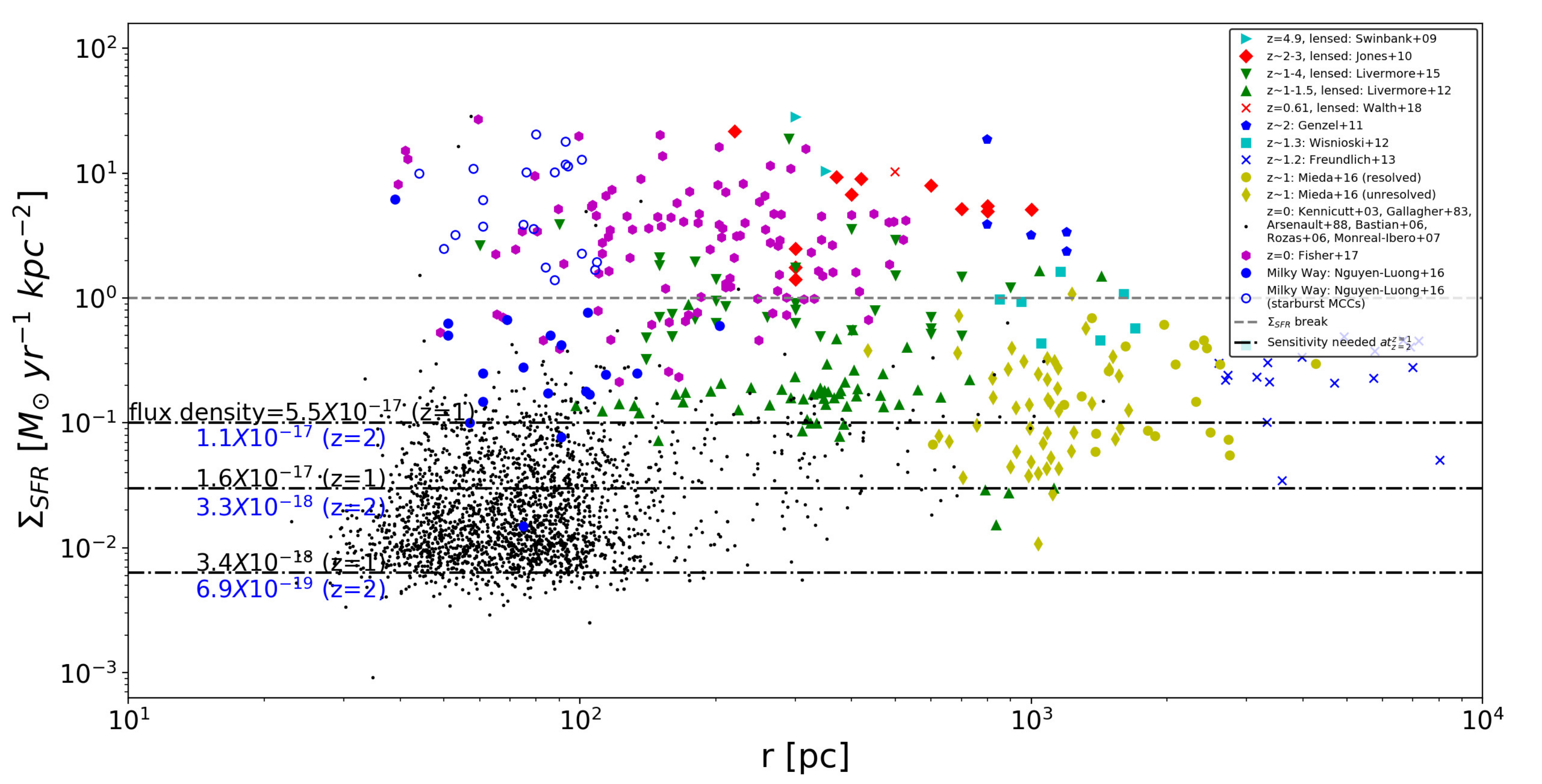}
\caption{Figure \ref{fig:SFRD} comparing clump size and \sfrd \, with additional curves related to the potential sensitivity limit. The dashed black lines show the observed flux density at that \sfrd \, for a $z=1$ (black text) and $z=2$ (blue text) source. All flux densities are in units of $\rm erg \ s^{-1} \ cm^{-2} \ arcsec^{-2}$. \label{fig:sensitivity}}
\end{figure*}

\clearpage

\section{Dynamical Mass of Clumps} \label{sec:dyn_mass}
The dynamical mass ($M_{dyn}$) was estimated for all data which included a measurement of the clump velocity dispersion ($\sigma$). This was calculated from Equation \ref{eqn:Mdyn} in order to estimate the dynamical time ($\tau_{dyn}$) of the clumps (Equation \ref{eqn:t_dyn}). 

\begin{eqnarray}
M_{dyn} = 5 \frac{\sigma^2 r_{clump}}{G} \label{eqn:Mdyn}\\
\tau_{dyn}=\sqrt{\frac{3\pi}{32G\rho}}, \ \ \rho=\frac{M_{dyn}}{\frac{4}{3}\pi r_{clump}^3} \label{eqn:t_dyn}
\end{eqnarray}

The average $M_{dyn}$ of all clumps with measured velocity dispersion is $2.7\times10^9$\,\msun \,  with an average dynamical time of 3.5\,Myr. The results of these calculations are listed in Table \ref{tbl:dyn_mass} for the IROCKS \citep{mied16} clumps as a sample.

\setcounter{table}{0}

\startlongtable
\begin{deluxetable}{lcc}
\tablewidth{0.5\textwidth}
\tabletypesize{\small}
\tablecaption{Dynamical Mass and Time Estimates for IROCKS sources \label{tbl:dyn_mass}}
\tablehead{
\colhead{Object} & \colhead{$M_{dyn}$} & \colhead{$\tau_{dyn}$} \\
\colhead{(Galaxy-Clump)} & \colhead{$\rm(10^{9} $\,\msun)}& \colhead{$\rm(10^6 \ yrs)$}
}
\tablecolumns{3}
\startdata
\tableline
\multicolumn{3}{c}{Resolved} \\
\tableline
32016379-1 & 0.52 & 8.11 \\
32016379-2 & 6.79 & 7.8 \\
32036760-0 & 8.79 & 16.48 \\
33009979-0 & 6.95 & 9.69 \\
33009979-1 & 2.75 & 11.0 \\
42042481-4 & 11.8 & 16.3 \\
DEEP11026194-1 & 11.36 & 12.97 \\
DEEP12008898-0 & 9.45 & 14.01 \\
DEEP12008898-1 & 10.55 & 14.33 \\
DEEP12008898-2 & 5.03 & 10.03 \\
DEEP12019627-0 & 3.97 & 16.07 \\
DEEP12019627-2 & 2.86 & 9.45 \\
DEEP13017973-3 & 3.73 & 24.67 \\
DEEP13017973-6 & 41.19 & 3.07 \\
TKRS11169-2 & 21.39 & 7.4 \\
TKRS11169-3 & 39.45 & 8.39 \\
TKRS7187-4 & 24.59 & 11.44 \\
TKRS7615-01-1 & 6.67 & 7.86 \\
TKRS7615-01-2 & 13.26 & 8.31 \\
TKRS9727-4 & 39.91 & 17.36 \\
UDS11655-0 & 5.58 & 15.84 \\
\tableline
\multicolumn{3}{c}{Unresolved} \\
\tableline
32016379-0 & 4.23 & 6.6 \\
32040603-0 & 2.64 & 5.7 \\
33009979-2 & 4.09 & 7.11 \\
42042481-0 & 5.15 & 3.24 \\
42042481-1 & 3.49 & 8.18 \\
42042481-2 & 1.21 & 11.01 \\
42042481-3 & 2.44 & 9.23 \\
42042481-5 & 13.27 & 1.82 \\
42042481-6 & 2.9 & 7.26 \\
DEEP11026194-0 & 10.61 & 7.46 \\
DEEP12019627-1 & 3.17 & 6.19 \\
DEEP12019627-3 & 5.4 & 8.49 \\
DEEP12019627-4 & 0.81 & 15.58 \\
DEEP12019627-5 & 6.77 & 5.9 \\
DEEP13017973-0 & 1.98 & 9.98 \\
DEEP13017973-1 & 1.84 & 2.62 \\
DEEP13017973-2 & 5.9 & 7.77 \\
DEEP13017973-4 & 6.15 & 4.3 \\
DEEP13017973-5 & 11.09 & 2.14 \\
DEEP13017973-7 & 2.61 & 8.08 \\
DEEP13043023-0 & 2.74 & 4.27 \\
DEEP13043023-1 & 6.38 & 6.03 \\
DEEP13043023-2 & 18.88 & 5.19 \\
DEEP13043023-3 & 4.96 & 6.82 \\
J033249.73-0 & 6.97 & 9.28 \\
J033249.73-1 & 7.66 & 2.95 \\
J033249.73-2 & 4.32 & 2.95 \\
J033249.73-3 & 5.19 & 7.39 \\
TKRS11169-0 & 5.16 & 4.57 \\
TKRS11169-1 & 4.97 & 5.27 \\
TKRS11169-4 & 5.22 & 6.14 \\
TKRS7187-0 & 4.02 & 6.95 \\
TKRS7187-2 & 6.99 & 5.52 \\
TKRS7187-3 & 0.72 & 15.37 \\
TKRS7187-5 & 10.34 & 3.29 \\
TKRS7187-6 & 5.52 & 4.71 \\
TKRS7615-01-0 & 5.16 & 7.57 \\
TKRS7615-01-3 & 12.09 & 7.51 \\
TKRS7615-01-4 & 4.24 & 5.97 \\
TKRS7615-01-5 & 8.93 & 7.89 \\
TKRS9727-0 & 2.08 & 9.77 \\
TKRS9727-1 & 13.22 & 6.38 \\
TKRS9727-3 & 2.01 & 6.51 \\
TKRS9727-5 & 2.69 & 9.05 \\
UDS10633-0 & 4.68 & 7.81 \\
UDS11655-1 & 5.28 & 5.32
\enddata
\tablecomments{Not all data sets included measurements of $\sigma_{clump}$.}
\end{deluxetable}

\end{document}